\documentclass[aps,prd,reprint,showpacs,showkeys,amsmath,amssymb,eqsecnum,nofootinbib,notitlepage,superscriptaddress,floatfix,twocolumn,longbibliography]{revtex4-1}

\usepackage{hyperref}
\usepackage{dcolumn}	
\usepackage{longtable}
\usepackage{graphicx}
\usepackage{url}
\usepackage{bm}		
\usepackage{bbm}
\usepackage{color}
\usepackage[usenames,dvipsnames,svgnames,table]{xcolor}

\newcommand{\etal}{{\it et~al.\hbox{}\/}}
\newcommand{\Subroutine}[1]{\textsf{#1}}
\newcommand{\SubroutineLibrary}[1]{\textsf{#1}}
\newcommand{\mathematica}{{\scshape Mathematica}}

\newcommand{\xcsy}[2]{{\setbox0=\hbox{#2}\hbox to \wd0{\hss{#1}\hss}}}

\newcommand{\boxop}{\Box}

\newcommand{\gtsim}{\gtrsim}
\newcommand{\del}{\nabla}
\renewcommand{\O}{\mathcal{O}}
\newcommand{\Scri}{\mathcal{J}}
\newcommand{\sun}{\odot}
\newcommand{\realpart}[1]{\mathop{\text{Re}}\left(#1\right)}
\newcommand{\imagpart}[1]{\mathop{\text{Im}}\left(#1\right)}

\newcommand{\ZZ}{\mathbbm{Z}}					

\newcommand{\ssl}{\mathfrak{l}} 

\newcommand{\abs}[1]{\lvert#1\rvert}			

\newcommand{\constant}{\text{constant}}
\newcommand{\fit}{\text{fit}}
\newcommand{\numerical}{\text{numerical}}
\newcommand{\observer}{\text{observer}}
\newcommand{\particle}{\text{particle}}
\newcommand{\periapsis}{\text{periapsis}}
\newcommand{\physical}{\text{physical}}
\newcommand{\puncture}{\text{puncture}}

\newcommand{\rref}{\text{ref}}
\newcommand{\regularized}{\text{regularized}}

\newcommand{\watchpoint}{\text{watchpoint}}

\newcommand{\hh}[1]{\left(#1\right) }			
\newcommand{\sqb}[1]{\left[#1\right]}			

\newcommand{\tst}{\rule[-9pt]{0pt}{24pt}}		

\newcommand{\eg}{e.g.\hbox{}}
\newcommand{\nb}{n.b.\hbox{}}

\newcommand{\WigglesNo}{\textcolor{red}{$\bm{\times}$}}

\newcommand{\WigglesPossible}{\textcolor{Tan}{$\bm{\diamondsuit}$}}

\newcommand{\WigglesLikely}{\textcolor{blue}{$\bm{\sim}$}}

\newcommand{\WigglesYes}{\textcolor{green}{$\bm{+}$}}

\newcommand{\WigglesYesMC}{\textcolor{green}{$\bm{\oplus}$}}

\begin{document}
\title{Excitation of {K}err quasinormal modes in extreme--mass-ratio inspirals}

\author{Jonathan Thornburg} 
\affiliation{Department of Astronomy and Center for Spacetime Symmetries,
             Indiana University, Bloomington, Indiana 47405,
             USA}
\affiliation{Max-Planck-Institut f\"{u}r Gravitationsphysik,
             Albert-Einstein-Institut,
             Am M\"{u}hlenberg 1, D-14476 Potsdam-Golm,
             Germany}
\author{Barry Wardell}
\affiliation{School of Mathematics and Statistics,
             University College Dublin, Belfield,
             Dublin 4, DO4 V1W8,
             Ireland}
\author{Maarten van de Meent} 
\affiliation{Max-Planck-Institut f\"{u}r Gravitationsphysik,
             Albert-Einstein-Institut,
             Am M\"{u}hlenberg 1, D-14476 Potsdam-Golm,
             Germany}
\affiliation{Mathematical Sciences and STAG Research Centre,
         	University of Southampton, Southampton, SO17 1BJ,
         	United Kingdom}

\date{\today}

\begin{abstract}
If a small compact object orbits a
black hole, it is known that it can excite the black hole's
quasinormal modes (QNMs), leading to high-frequency oscillations
(``wiggles'') in the radiated field at $\Scri^+$, and in the
radiation-reaction self-force acting on the object after its orbit
passes through periapsis.
Here we survey the phenomenology of these wiggles
across a range of black hole spins and equatorial orbits.
In both the scalar-field and gravitational cases we find that wiggles
are a generic feature across a wide range of parameter space,
and that they are observable in field perturbations at fixed spatial
positions, in the self-force, and in radiated fields at $\Scri^+$.
For a given charge or mass of the small body,
the QNM excitations have the highest
amplitudes for systems with a highly spinning central black hole, a
prograde orbit with high eccentricity, and an orbital periapsis
close to the light ring.
The QNM amplitudes depend smoothly on the orbital parameters, with
only very small amplitude changes when the orbit's (discrete) frequency
spectrum is tuned to match QNM frequencies.
The association of wiggles with QNM excitations suggest that they
represent a situation where the \emph{nonlocal} nature of the
self-force is particularly apparent, with the wiggles arising as
a result of QNM excitation by the compact object near periapsis,
and then encountered later in the orbit.
Astrophysically, the effects of wiggles at $\Scri^+$ might allow direct
observation of Kerr QNMs in extreme-mass-ratio inspiral (EMRI)
binary black hole systems, potentially enabling new tests of
general relativity.
\end{abstract}

\maketitle 


\section{Introduction}
\label{sect:introduction}

Consider a small (compact) body of mass $\mu M$ (with $0 < \mu \ll 1$)
moving freely near a Schwarzschild or Kerr black hole of mass~$M$.
This system emits gravitational radiation, and there is a corresponding
radiation-reaction influence on the small body's motion.  Calculating
the resulting perturbed spacetime (including the small body's motion
and the emitted gravitational radiation) is a long-standing research
question in general relativity.

There is also an astrophysical motivation for this calculation:
If a neutron star or stellar-mass black hole of mass~${\sim}\, 1$--$100 M_\sun$
orbits a massive black hole of mass~${\sim}\, 10^5$--$10^7 M_\sun$,
\footnote{
	 $M_\sun$ denotes the solar mass.
	 }
{} the resulting ``extreme--mass-ratio inspiral'' (EMRI) system is
expected to be a strong astrophysical gravitational-wave (GW) source
detectable by the planned Laser Interferometer Space Antenna (LISA)
space-based gravitational-wave detector.  LISA is expected to observe
many such systems, some of them at quite high signal-to-noise ratios
(\cite{Gair-etal-2004:LISA-EMRI-event-rates,Barack-Cutler-2004,
Amaro-Seoane-etal-2007:LISA-IMRI-and-EMRI-review,
Gair-2009:LISA-EMRI-event-rates}).
The data analysis for, and indeed the detection of, such systems
will generally require matched-filtering the detector data stream
against appropriate precomputed GW templates.  The problem of
computing such templates provides an astrophysical motivation for
EMRI modelling.

In the test-particle limit it has long been known that an unbound
(scattering) flyby can excite quasinormal modes (QNMs) of the background
black hole.
Kojima and Nakamura~\cite{Kojima-Nakamura-1984} studied this process,
finding that
``A scattered particle excites the quasi-normal mode under the condition
that twice the angular velocity at the periapsis is greater than the
real part of the frequency of the quasi-normal mode''.
Their figure~3(b) shows an example of the QNM oscillations in the
radiated gravitational waves at $\Scri^+$.

Burko and Khanna~\cite{Burko-Khanna-2007} found small oscillations
in the total radiated energy flux from a test particle making a parabolic
(unbound) flyby of a Kerr black hole.  They attributed these oscillations
to the particle encountering scattered gravitational waves emitted during
the particle's inbound motion.

O'Sullivan and Hughes~\cite{OSullivan:2015lni} observed
``small-amplitude high-frequency oscillations'' in their calculations of
the horizon shear of a Kerr black hole orbited by a test particle.
Because they did not find corresponding oscillations in the horizon's
tidal distortion field, and their measured oscillation frequencies did not
match known Kerr QNM frequencies, they concluded that the horizon-shear
oscillations they observed
``cannot be related to the [Kerr black] hole's quasi-normal modes''.

Thornburg and Wardell~\cite{Thornburg-Wardell-2017:Kerr-scalar-self-force}
(hereinafter TW) calculated the scalar-field self-force for eccentric
equatorial particle orbits in Kerr spacetime.  For some systems where
the Kerr black hole was highly spinning and the particle orbit was prograde
and highly eccentric, TW found that the self-force exhibits large oscillations
(``wiggles'') on the outgoing leg of the orbit shortly after periapsis
passage.  TW suggested that wiggles ``are in some way \emph{caused}
by the particle's close passage by the large black hole''.
Thornburg~\cite{Thornburg-2016-Capra-Meudon-talk,
Thornburg-2017-Capra-Chapel-Hill-talk}
presented fits of damped-sinusoid models
to these wiggles for a range of Kerr spins and particle orbits,
found close agreement of the model complex-frequencies with those of
known Kerr QNMs, and argued that this agreement shows that wiggles are,
in fact, caused by Kerr QNMs excited by the particle's close periapsis
passage.

Nasipak, Osburn,
and Evans~\cite{Nasipak-Osburn-Evans-2019:Kerr-scalar-self-force-and-wiggles}
calculated the scalar-field self-force and the radiated field at $\Scri^+$
for eccentric inclined particle orbits in Kerr spacetime.  For one particular
(equatorial) orbit configuration they confirmed TW's finding of wiggles
in the self-force and also found wiggles in the radiated scalar field
at $\Scri^+$, fitting these to a superposition of $\ell\,{=}\,m\,{=}\,1$,
$\ell\,{=}\,m\,{=}\,2$, $\ell\,{=}\,m\,{=}\,3$, and $\ell\,{=}\,m\,{=}\,4$
Kerr quasinormal modes (QNMs).  They concluded that wiggles are caused by
Kerr QNMs excited by the particle's close periapsis passage.

Rifat, Khanna,
and Burko~\cite{Rifat-Khanna-Burko-2019:wiggles-in-near-extremal-Kerr}
recently studied wiggles for EMRIs where the central Kerr BH is nearly
extremal (dimensionless spin up to $0.999\,99$), finding that in such
systems many Kerr QNMs can be simultaneously excited.

Here we extend Refs.~\cite{Thornburg-2016-Capra-Meudon-talk,
Thornburg-2017-Capra-Chapel-Hill-talk} and survey wiggles' phenomenology
over a wide range in
parameter space for eccentric equatorial orbits in Kerr spacetime,
for both the scalar-field model and the full gravitational field.
We focus on leading-order radiation and radiation-reaction effects
computed via 1st-order perturbations of Kerr spacetime,
i.e., (for the gravitational case)
$\O(\mu)$~field perturbations near the Kerr black hole,
$\O(\mu)$ radiation at $\Scri^+$,
and $\O(\mu^2)$~radiation-reaction ``self-forces'' acting on the small body.
We fit models of Kerr QNMs to all these diagnostics.

Our focus is the case where the perturbing body's orbit is highly
relativistic, so post-Newtonian methods (see, for
example,~\cite[Section~6.10]{Damour-in-Hawking-Israel-1987};
\cite{Poisson-Will-book-2014,Will-TEGP-2nd-Ed,
Blanchet-2014-living-review,Futamase-Itoh-2007:PN-review,
Blanchet-2009:PN-review,Schaefer-2009:PN-review} and references therein)
are not reliably accurate.  Since the timescale for radiation reaction
to shrink the orbit is very long ($\sim \mu^{-1} M$) while the required
resolution near the small body is very high ($\sim \mu M$), a direct
``numerical relativity'' integration of the Einstein equations
(see, for example,~\cite{Pretorius-2007:2BH-review,
Hannam-etal-2009:Samurai-project,Hannam-2009:2BH-review,
Hannam-Hawke-2010:2BH-in-era-of-Einstein-telescope-review,
Campanelli-etal-2010:2BH-numrel-review} and references therein)
would be prohibitively expensive (and probably insufficiently accurate)
for this problem.
\footnote{
	 A number of researchers have attempted direct
	 numerical-relativity binary black hole simulations
	 for systems with ``intermediate'' mass ratios up to
	 $100\,{:}\,1$ ($\mu = 0.01$), (see, for example,
\protect\cite{Bishop-etal-2003,Bishop-etal-2005,
Sopuerta-etal-2006,Sopuerta-Laguna-2006,
Lousto-etal-2010:intermediate-mass-2BH-numrel-Lazarus,
Lousto-Zlochower-2011:100-to-1-mass-ratio-2BH,
Husa-etal-2015:frequency-domain-GW-model-q-up-to-18}).
	 However, it has not (yet) been possible to extend
	 these results to the extreme-mass-ratio case
	 nor to accurately evolve any systems with mass
	 ratios more extreme than $18\,{:}\,1$
	 for a radiation-reaction
	 time scale~\cite{Husa-etal-2015:frequency-domain-GW-model-q-up-to-18}
	 (Hinder~\cite{Hinder-pers-comm-2019:32:1-2BH-in-progress}
	 reports ongoing efforts to extend this to~$32\,{:}\,1$).
	 }
{}  Instead, we use black hole perturbation theory, treating the
small body as an $\O(\mu)$~perturbation on the background spacetime.

We present results obtained from two separate numerical codes which
were programmed independently, using completely different theoretical
formalisms and numerical methods:
\begin{itemize}
\item	Our scalar-field results were obtained using TW's
	code~\cite{Wardell-etal-2012,
Thornburg-Wardell-2017:Kerr-scalar-self-force}
	extended to compute additional field diagnostics.
	This code uses an effective-source regularization followed by
	an $e^{im\phi}$ Fourier-mode decomposition and a separate
	$2{+}1$-dimensional time-domain numerical evolution of each
	Fourier mode.
	The main outputs of this code are the regularized scalar field
	at selected (fixed) spatial positions, the regularized scalar field
	and self-force at the particle, and the physical radiated
	scalar field at $\Scri^+$.
\item	Our gravitational-field results were obtained using
	van de Meent's code~\cite{vandeMeent:2015lxa,vandeMeent:2016pee,
vandeMeent:2016hel,vandeMeent:2017bcc}.
	This code obtains the local metric perturbation in the
	frequency domain by reconstructing the metric perturbation
	from the Weyl scalar  $\Psi_4$, followed by $\ell$-mode
	regularization to obtain the regular part.
	The main outputs of this code are the regularized
	outgoing--radiation-gauge metric perturbation and self-force
	at the particle, and the physical radiated $\Psi_4$ at $\Scri^+$
	and selected fixed positions in the spacetime.
\end{itemize}
For both codes we take the particle orbit to be a bound equatorial
geodesic; we do not consider changes in the orbit induced by the
self-force.

To briefly summarize our main results, we observe wiggles across a
wide range of Kerr spins and particle orbits.  Wiggles are present in
all of our field diagnostics in the strong-field region and at $\Scri^+$.
Except for a few anomalous results for near-extremal Kerr spacetimes
(dimensionless Kerr spins $\gtsim 0.9999$), our results are
all consistent with the interpretation of wiggles as Kerr QNMs.
Wiggles are stronger and more readily observable for high Kerr spins,
highly eccentric prograde particle orbits, and particle periapsis radii
close to the light ring.

The remainder of this paper is organized as follows:

Section~\ref{sect:introduction/notation-etal} summarizes our notation
and our parameterization of bound geodesic orbits in Kerr spacetime.
Section~\ref{sect:calculations-of-Kerr-perturbations}
briefly summarizes our calculations of
scalar-field~(section~\ref{sect:calculations-of-Kerr-perturbations/scalar-field})
and
gravitational~(section~\ref{sect:calculations-of-Kerr-perturbations/gravitational})
perturbations of Kerr spacetime.
Section~\ref{sect:field-diagnostics} describes our local-
and radiated-field diagnostics.
Section~\ref{sect:QNM-models-and-fits} describes our QNM models and
how we fit these to time series of our field diagnostics.
Section~\ref{sect:data-and-QNM-fits} presents our data for
scalar-field and gravitational perturbations of Kerr spacetime, and QNM-model
fits to this data.
Finally, section~\ref{sect:discussion-and-conclusions} discusses our
results and presents our conclusions.


\subsection{Notation and parameterization of {K}err geodesics}
\label{sect:introduction/notation-etal}

We generally follow the sign and notation conventions of Wald~\cite{Wald-1984},
with $G = c = 1$ units and a $(-,+,+,+)$ metric signature.  We use the
Penrose abstract-index notation, with indices $abcd$ running over
spacetime coordinates, and $ijk$ running over the spatial coordinates.
$\del_a$ is the (spacetime) covariant derivative operator
and $g$ is the determinant of the spacetime metric.
$X := Y$ means that $X$ is defined to be $Y$.
$\boxop := \del_a \del^a$ is the 4-dimensional (scalar) wave
operator~\cite{Brill-etal-1972,Teukolsky73}.

We use Boyer-Lindquist coordinates $(t,r,\theta,\phi)$ on Kerr spacetime,
defined by the line element
\begin{align}
ds^2	= {} &
		- \left( 1 - \frac{2Mr}{\Sigma} \right) \, dt^2
		- 4M^2 \tilde{a} \frac{r \sin^2\theta}{\Sigma} \, dt \, d\phi
								\nonumber\\*
	&
		+ \frac{\Sigma}{\Delta} \, dr^2
		+ \Sigma \, d\theta^2
								\nonumber\\*
	&
		+ \left(
		  r^2 + M^2 \tilde{a}^2 + 2M^3 \tilde{a}^2 \frac{r \sin^2\theta}{\Sigma}
		  \right) \sin^2\theta \, d\phi^2
								\text{~,}
					 \label{eqn:Kerr-Boyer-Lindquist-coords}
\end{align}
where $M$~is the black hole's mass,
$\tilde{a} = J/M^2$ is the black hole's dimensionless spin
(limited to $|\tilde{a}| < 1$),
$\Sigma := r^2 + M^2 \tilde{a}^2 \cos^2\theta$,
and
$\Delta := r^2 - 2Mr + M^2 \tilde{a}^2$.
We also define $a := M\tilde{a}$
(this is unrelated to the use of $a$ as an abstract tensor index).
In these coordinates the event horizon is the coordinate sphere
$r = r_+ = M \left(1 + \sqrt{1 - \tilde{a}^2}\right)$
and the inner horizon is the coordinate sphere
$r = r_- = M \left(1 - \sqrt{1 - \tilde{a}^2}\right)$.
(See footnote~\ref{footnote:sf-compactification} for a discussion
of TW's coordinate compactification near the event horizon and $\Scri^+$.)

Following Ref.~\cite{Sundararajan-Khanna-Hughes-2007},
we define the tortoise coordinate~$r_*$ by
\begin{equation}
\frac{dr_*}{dr} = \frac{r^2+M^2 \tilde{a}^2}{\Delta}.
							  \label{eqn:rstar-defn}
\end{equation}
and fix the additive constant by choosing
\begin{align}
r_*	= {}&	r
		+ 2M\frac{r_+}{r_+ - r_-} \ln\left( \frac{r - r_+}{2M} \right)
								\nonumber\\
	&	\phantom{r}
		- 2M\frac{r_-}{r_+ - r_-} \ln\left( \frac{r - r_-}{2M} \right)
								\text{~.}
							    \label{eqn:rstar(r)}
\end{align}
$u := t - r_*$ is thus an outgoing null coordinate.

The particle's worldline is $x^i = x^i_\particle(t)$; we consider
this to be known in advance, i.e., we do \emph{not} consider changes to
the particle's worldline induced by the self-force.
For present purposes we consider only particle orbits in the Kerr
spacetime's equatorial plane; this restriction is for computational
convenience and is not fundamental.
We take the particle to orbit in the $d\phi/dt > 0$ direction,
with $\tilde{a} > 0$ for prograde orbits and $\tilde{a} < 0$
for retrograde orbits.

We define $r_{\min}$ and $r_{\max}$ to be the particle's periapsis and
apoapsis $r$~coordinates, respectively.  We parameterize bound geodesic
equatorial particle orbits by the usual (dimensionless) semi-latus rectum~$p$
and eccentricity~$e$
\big(defined by $r_{\min} = pM\big/(1+e)$ and $r_{\max} = pM\big/(1-e)$\big),
so that the particle orbit is given by
\begin{equation}
r_\particle(t) = \frac{pM}{1 + e \cos \chi_r(t)}
								\text{~,}
\end{equation}
for a suitable orbital-phase function $\chi_r$.
We refer to the combination of a spacetime and a (bound geodesic)
particle orbit as a ``configuration'', and parameterize it with the
triplet $(\tilde{a},p,e)$.
We define $T_r$ to be the coordinate-time period of the particle's
radial motion; we usually refer to $T_r$ as the particle's ``orbital period''.



\section{Calculations of scalar-field and gravitational perturbations
         of {K}err spacetime}
\label{sect:calculations-of-Kerr-perturbations}


\subsection{Scalar-field perturbations of {K}err spacetime}
\label{sect:calculations-of-Kerr-perturbations/scalar-field}

In this section we summarize TW's scalar-field
calculations~\cite{Wardell-etal-2012,
Thornburg-Wardell-2017:Kerr-scalar-self-force}.
These authors consider a real scalar field~$\Phi_\physical$ satisfying
the wave equation in Kerr spacetime,
\begin{equation}
\boxop \Phi_\physical
	= -4 \pi q
	  \int \frac{\delta^4 \bigl( x^a - x^a_\particle(t) \bigr)}{\sqrt{-g}}
	       \, d\tau
					       \label{eqn:scalar-field-wave-eqn}
								\text{~,}
\end{equation}
sourced by a point ``particle'' of scalar charge~$q$ which is taken to
move on a (pre-specified) equatorial geodesic orbit.  $\Phi_\physical$
satisfies outgoing-radiation (retarded) boundary conditions at $\Scri^+$.
This system provides a toy model of the full gravitational perturbation
problem without the complexity of gauge choice.

Because $\Phi_\physical$ diverges on the particle worldline,
\eqref{eqn:scalar-field-wave-eqn} must be regularized.
TW use an effective-source regularization of the type introduced
by Barack and Golbourn~\cite{Barack-Golbourn-2007} and
Vega and Detweiler~\cite{Vega-Detweiler-2008:self-force-regularization}
(see~\cite{Vega-Wardell-Diener-2011:effective-source-for-self-force}
for a review), using the puncture function and effective source
described by Wardell~\etal~\cite{Wardell-etal-2012}.  In a neighborhood
of the particle worldline, TW define a (real) regularized scalar field
$\Phi_\regularized = \Phi_\physical - \Phi_\puncture$,
where $\Phi_\puncture$ is a suitably-chosen approximation to the
Detweiler-Whiting singular field~\cite{Detweiler-Whiting-2003}.
The (4-vector) self-force acting on the particle is then given by
\begin{equation}
F_a = \bigl. (\del_a \Phi_\regularized) \bigr| _{x^i = x^i_\particle(t)}
					     \label{eqn:scalar-field-self-force}
								\text{~.}
\end{equation}

TW make an azimuthal Fourier decompositions of all the spacetime
scalar fields into complex $e^{im\tilde{\phi}}$~modes,
\begin{equation}
\Phi(t,r,\theta,\phi)
	= \sum_{m=-\infty}^\infty
	  \dfrac{1}{r} \varphi_m(t,r,\theta) e^{im\tilde{\phi}} 
					       \label{eqn:scalar-field-mode-sum}
								\text{~,}
\end{equation}
where the extra factor of $1/r$~is introduced for computational convenience
and where $\tilde{\phi} := \phi + f(r)$ is an ``untwisted'' azimuthal
coordinate, with
\begin{equation}
f(r) = \frac{\tilde{a}}{2\sqrt{1-\tilde{a}^2}}
       \ln \left| \frac{r-r_+}{r-r_-} \right|
								\text{~.}
\end{equation}

For each~$m$-mode, TW introduce a finite worldtube surrounding the
particle worldline in $(t,r,\theta)$ space.  For particle orbits with
significant eccentricity ($e \gtsim 0.2$) the worldtube (now viewed
as a region in $(r,\theta)$ in each $t=\constant$ slice) moves inward
and outward in $r$ during each orbit so as to always contain the particle.
All the results reported here were obtained using a worldtube which
is rectangular in~$(r,\theta)$, with size~$10M$ in~$r_*$ by
approximately $0.25$~radians in $\theta$, symmetric about the
equatorial plane, and kept centered on the particle to within~$0.25M$
in~$r_*$ as the particle moves.

TW numerically solve for the piecewise function
\begin{equation}
\bigl. (\varphi_m) \bigr. _\numerical
	= \begin{cases}
	  \bigl. (\varphi_m) \bigr. _\regularized
			& \text{inside the worldtube}		\\
	  \bigl. (\varphi_m) \bigr. _\physical
			& \text{outside the worldtube}		
	  \end{cases}
								\text{~.}
\end{equation}
using a time-domain $2{+}1$-dimensional finite-difference
numerical evolution with mesh refinement.  
Because the (Kerr) background spacetime is axisymmetric, the Fourier
modes~$\varphi_m$ evolve independently -- there is no mixing of the modes.
Because the physical scalar field $\Phi$ is real, only the $m \ge 0$~modes
need to be explicitly computed; the $m < 0$~modes may be obtained by symmetry.

Corresponding to the Fourier decomposition~\eqref{eqn:scalar-field-mode-sum},
the self-force~\eqref{eqn:scalar-field-self-force}
can be written as a similar sum of $e^{im\tilde{\phi}}$~modes,
\begin{equation}
F_a = q \sum_{m=0}^\infty F_a^{(m)}
						 \label{eqn:scalar-self-force-mode-sum}
								\text{~,}
\end{equation}
where each $F_a^{(m)}$ may be computed from the corresponding
$\bigl. (\varphi_m) \bigr. _\regularized$ field near the particle.
TW compute a finite set of modes (typically $0 \le m \le 20$) and
estimate the $m > 20$ contributions to the
sum~\eqref{eqn:scalar-self-force-mode-sum}
via a large-$m$ asymptotic series.

TW use a Zengino\u{g}lu-type hyperboloidal compactification~
\cite{Zenginoglu-2008:hyperboloidal-foliations-and-scri-fixing,
Zenginoglu-2008:hyperboloidal-evolution-with-Einstein-eqns,
Zenginoglu-2011:hyperboloidal-layers-j-comp-phys,
Zenginoglu-Khanna-2011:Kerr-EMRI-waveforms-via-Teukolsky-evolution,
Zenginoglu-Kidder-2010:hyperboloidal-evolution-of-scalar-field-on-Schw,
Zenginoglu-Tiglio-2009:spacelike-matching-to-null-infinity,
Bernuzzi-Nagar-Zenginoglu-2011:Schw-EMRI-waveforms-via-EOB-evolution,
Bernuzzi-Nagar-Zenginoglu-2012:Schw-EMRI-horizon-absorption-effects}
so they also have direct access to far-field quantities at $\Scri^+$
(where the coordinate~$t$ becomes a Bondi-type retarded time coordinate).
\footnote{
\label{footnote:sf-compactification}
	 More precisely, TW define compactified coordinates~$(T,R_*)$
	 which are identical to (respectively) the Boyer-Lindquist~$t$
	 and the tortoise coordinate~$r_*$ throughout a neighborhood
	 of the region $r_{\min} \le r \le r_{\max}$ containing the
	 particle orbit, but which are compactified near the event
	 horizon and $\Scri^+$.
	 $T$~is a Bondi-type retarded time coordinate at $\Scri^+$.
	 For present purposes the details of the compactification
	 are not important, so for convenience of exposition we
	 refer to $T$ as $t$ hereinafter when describing our
	 diagnostics at $\Scri^+$.
	 }


\subsection{Gravitational perturbations of {K}err spacetime}
\label{sect:calculations-of-Kerr-perturbations/gravitational}

In this section we summarize the metric reconstruction approach
used by van de Meent~\cite{vandeMeent:2015lxa,vandeMeent:2016pee,
vandeMeent:2016hel,vandeMeent:2017bcc} to calculate
gravitational perturbation of Kerr spacetime generated by particles
on bound geodesic orbits.
This approach starts from the the spin-(-2) Teukolsky variable,
\begin{equation}
\Phi_{-2} := (r-i a\cos\theta)^4\Psi_4,
\end{equation}
where $\Psi_4$ is one of the Weyl scalars. As shown by
Teukolsky~\cite{Teukolsky73}, linear perturbations to this variable satisfy
an equation of motion that decouples from all other degrees of freedom.
Moreover, unlike the linearized Einstein equation, solutions to the
Teukolsky equation can be decomposed into Fourier-harmonic modes,
\begin{equation}
\Phi_{-2} = \sum_{\ssl m \omega}
	    R_{\ssl m \omega}(r) S_{\ssl m\omega}(\theta)
	    e^{i m \phi- i \omega t}
								\text{~,}
\end{equation}
where
the $R_{\ssl m \omega}$ are solutions of the radial Teukolsky equation,
the $S_{\ssl m\omega}$ are spin-weight spheroidal harmonics,
and $\ssl$ is the spheroidal mode number.
In van de Meent's code the radial Teukolsky equation can be
solved to arbitrarily high precision using a numerical implementation
of the semi-analytical methods of
Mano, Suzuki, and Takasugi~\cite{Mano:1996gn,Mano:1996vt}. 

Remarkably, $\Phi_{-2}$ contains almost all information about the
corresponding perturbation of the metric~\cite{Wald:1973}, and in vacuum
it is possible to reconstruct the metric perturbation in a radiation
gauge~\cite{Cohen:1974cm,Kegeles:1979an,Chrzanowski:1975wv,Wald:1978vm}.
As detailed in Refs.~\cite{vandeMeent:2015lxa,vandeMeent:2016pee,
vandeMeent:2016hel,vandeMeent:2017bcc},
this procedure can be used to calculate the backreaction of the
metric perturbation on the particle, the gravitational self-force,
which then takes the form
\begin{equation} F_a
	= \sum_{\substack{\ssl m \omega\\ nk\pm}}
	  \mathcal{C}^{\pm}_{a m \omega nk}
	  R_{\ssl m \omega}^{(n)\pm}(r_0)
	  S_{\ssl m\omega}^{(k)}(\theta_0)
	  e^{i m \phi_0 - i \omega t_0}
								\text{~,}
\end{equation}
where the $ R_{\ssl m \omega}^{\pm}(r_0)$
are vacuum solutions of the radial Teukolsky equation analytically
extended to the particle position $r_0$ from either infinity ($+$)
or the black hole horizon ($-$) (method of extended homogeneous solutions~\cite{Barack:2008ms}), and the $(n)$ and $(k)$ superscripts on a function denote derivatives with respect to the argument. The indices $n$ and $k$ run from 0 to 3.

Although each individual term in
the sum above is finite, the sum as a whole does not converge. This
is a simple consequence of the fact that it was built from the
retarded field perturbation rather than the Detweiler-Whiting regular
field. To obtain the regular field we still need to subtract the
Detweiler-Whiting singular field. In principle this can be done mode-by-mode.
To match previous analytical calculations of the large $\ell$-behavior
of the singular field~\cite{Barack:2002mh,Barack:2009ux}, we need
to re-expand the spheroidal $\ssl$-modes to spherical $\ell$-modes,
\begin{equation}
F_a^{(\ell)} =
	\sum_{\substack{\ssl m \omega\\ n\pm}}
	\tilde{\mathcal{C}}^{\ell \pm}_{a\ssl m \omega n}
	R_{\ssl m \omega}^{(n)\pm}(r_0)
	Y_{\ell m}(\theta_0)
	e^{i m \phi_0 - i \omega t_0}
								\text{~.}
\end{equation}
With this re-expansion, the local gravitational
self force is given by
\begin{equation} F_a = \sum_\ell F_a^{(\ell)} - B_a
								\text{~,}
\end{equation}
where, as shown in~\cite{Pound:2013faa}, we can use the Lorenz-gauge~$B_a$
parameter given in~\cite{Barack:2002mh,Barack:2009ux}.

The metric reconstruction formalism can only recover parts of the metric
that contribute to $\Psi_4$. This means that the reconstructed metric
carries an ambiguity, which can be shown~\cite{Wald:1973} to consist of
perturbations of the background within the class of Kerr metrics
and pure gauge contributions. These ambiguities can be uniquely fixed
based on physical considerations~\cite{Merlin:2016boc,vandeMeent:2017fqk}.
The corrections needed to fix these ambiguities are known and straightforward
to add. They contribute only to the low frequency envelope of the
self-force. Hence, to facilitate identification and extraction of the
wiggles in the gravitational self-force, we omit them in this work.

Frequency domain calculations of the type used here are ideally
suited for calculating metric perturbations with a sparse discrete
frequency spectrum, such as those sourced by a particle on a low
eccentricity geodesic.  That spectrum becomes denser at higher
eccentricities, necessitating the calculation of more frequency modes and
making the calculation more time-consuming.  Moreover, as discussed
in detail in~Ref.~\cite{vandeMeent:2016pee}, the method of extended
homogeneous solutions leads to large cancellations between different
(low) frequency modes for high-$\ell$ modes, causing a large loss of
precision.  In this work we tackle this problem by harnessing the
full power of the arbitrary precision implementation of our code and
simply throw more precision at the computation than we lose in the
cancellation.

For this work we calculated the full gravitational self-force for
orbits with eccentricities up to $e = 0.8$, which involves dealing
with cancellations of around 30~orders of magnitude.  These calculations
are fairly resource intensive, taking $\O(10^4)$~CPU hours
(or a few days on 400~CPUs) to compute.

However, for many aspects of our investigation here, we do not need the
full local regular metric perturbation. If we want to look for the
dominant low-$\ssl$ QNMs, then these will contribute (mostly) to the
low-$\ssl$ modes of the gravitational metric perturbation.  For this
purpose, we define the individual~$\ssl m$ modes of the Teukolsky
variable
\begin{equation}
\Phi_{-2}^{(\ssl m)}
	= \sum_{\omega}
	  R_{\ssl m \omega}(r) S_{\ssl m\omega}(\theta)
	  e^{i m \phi - i \omega t},
\end{equation}
and the gravitational self-force
\begin{equation}
F_a^{(\ssl m)} :=
	\sum_{\substack{\omega\\ nk\pm}}
	\mathcal{C}^{\pm}_{a m \omega n k} R_{\ssl m \omega}^{(n)\pm}(r_0)
	S_{\ssl m\omega}^{(k)}(\theta_0) e^{i m \phi_0 - i \omega t_0}.
\end{equation}
These are much easier to compute, and for low~$\ssl$ do not suffer
from the large loss of precision due to the method of extended
homogeneous solutions, thus allowing for very high accuracy calculations
without excessive computational cost.


\section{Field diagnostics}
\label{sect:field-diagnostics}

We consider several different types of
local- and radiated-field diagnostics, and attempt to fit the wiggles
in these diagnostics to QNM models.  Clearly the presence of wiggles
in the physical scalar field or metric perturbation implies the presence
of wiggles in some or all of the $e^{im\tilde{\phi}}$~scalar-field modes
or $(\ssl m)$~metric-perturbation modes (respectively), and
vice versa.
\footnote{
	 While it would be theoretically possible for multiple
	 modes to have wiggles of the same frequency whose
	 amplitudes sum to approximately zero (leading to
	 an absence of wiggles in the physical fields), in
	 practice we have never observed this.
	 }
{}  Because many fewer QNMs are present at significant amplitude
(usually only one, or in a few cases two), it is much simpler to
analyze wiggles in the individual modes.

Table~\ref{tab:field-diagnostics} summarizes our local- and radiated-field
diagnostics for studying wiggles.  We consider (time series of) diagnostics
at three locations in spacetime:
\begin{itemize}
\item	\emph{Diagnostics of the local field at selected fixed spatial
	``watchpoint'' coordinate positions $(r,\theta,\phi) = \constant$.}
	These diagnostics directly sample any QNMs that may be
	present, but the diagnostics may be contaminated by the
	direct field when the particle is close to the watchpoint
	position.

	For the scalar-field case, we avoid any such possible
	contamination by considering the regularized field mode
	$\bigl. (\varphi_m) \bigr. _\regularized$.  However, this
	is only defined within the worldtube, so for orbits with
	significant eccentricity (where the worldtube moves in
	$(r,\theta)$ during the particle orbit) any given watchpoint
	may lie outside the worldtube (and thus leave
	$\bigl. (\varphi_m) \bigr. _\regularized$ undefined)
	during some parts of the orbit.  To minimize this effect,
	for many of the analyses reported here we use watchpoint
	positions which are near the orbit's apoapsis, where the
	particle (and hence the worldtube) motion is relatively slow
	and hence a suitable watchpoint can remain within the worldtube
	for a relatively long time in each orbit.
	All our scalar-field watchpoints are in the equatorial plane.


	For the gravitational case, the regularized field is not
	readily available, so instead we have the code
	output the retarded $\Phi_{-2}^{(\ssl m)}$ on the symmetry axis
	of the background Kerr spacetime ($\theta=0$)
	and the equatorial plane $(\theta=\pi/2)$ at coordinate radii
	corresponding to the particle's periapsis and apoapsis distances.

\item	\emph{Diagnostics of the local field at the particle.}
	Here we consider the $e^{im\tilde{\phi}}$ (scalar-field)
	and $(\ssl m)$ (gravitational) modes of the
	self-force itself.  The main complication here is that
	these diagnostics sample the field perturbation at a
	\emph{time-dependent} position (the particle position),
	so our fitting model for the QNM effects must include corrections
	for the spatial variation of the QNM eigenfunctions
	as the particle (sampling point) moves.
	For the azimuthal~($\phi$) particle motion this is
	straightforward
	(described in section~\ref{sect:QNM-models-and-fits})
	but for the radial~($r$) motion we include this
	correction only approximately.

\item \emph{Diagnostics of the radiated field at $\Scri^+$.}
	These have the advantage of being physically observable
	and of allowing the $e^{im\tilde{\phi}}$ (scalar-field)
        and $(\ssl m)$ (gravitational) mode decompositions to be
	defined in a weak-field region (for the gravitational case,
	this also avoids any gauge dependence).

	At $\Scri^+$ we only have the physical (retarded) fields available,
	so it is more difficult to separate wiggles from the overall
	radiation pattern.  To help in making this separation,
	we observe that wiggles are of relatively high (temporal)
	frequency relative to other major features in the
	radiated fields, so that taking time derivatives of the
	radiated fields increases the wiggles' amplitude relative
	to that of the other features.
	We have found that good results are obtained by using
	as diagnostics the second time derivatives
	$\partial_{tt} \left( \bigl. (\varphi_m) \bigr. _\physical \right)$
	evaluated in the equatorial plane (scalar field)
\footnote{
	 Recall (cf.~footnote~\protect\ref{footnote:sf-compactification})
	 that in our scalar-field computations, $t$ is a
	 Bondi-type retarded time coordinate at $\Scri^+$.
	 }
{}	and $\Psi_4$ evaluated either on the z~axis or
	in the equatorial plane (gravitational).
\end{itemize}

\begin{table}[t]
\begin{center}
\caption{
\label{tab:field-diagnostics}
	This table shows the local- and radiated-field diagnostics
	in which we study wiggles.
	}
\begin{ruledtabular}
\begin{tabular}{lll}
Field perturbation at \dots
			& Scalar field		& Gravitational		\\
\hline 
fixed spatial position
			& $\bigl. (\varphi_m) \bigr. _\regularized$
						& $\Phi_{-2}^{(\ssl m)}$	\\
(strong-field)		&			&			\\
\hline 
particle position 	& $F_a^{(m)}$		& ${}_{-2}F_a^{(\ssl m)}$
									\\
\hline 
$\Scri^+$		& $\partial_{tt}
			  \left( \bigl. (\varphi_m) \bigr. _\physical \right)$
						& $\Psi_4$		
\end{tabular}
\end{ruledtabular}
\end{center}
\end{table}


\section{Quasinormal-mode models and fits}
\label{sect:QNM-models-and-fits}


\subsection{Scalar-field perturbations}
\label{sect:QNM-models-and-fits/scalar-field}


\subsubsection{Perturbations at a fixed spatial position}
\label{sect:QNM-models-and-fits/scalar-field/watchpoint}

Consider first the case of wiggles in an individual $e^{im\phi}$~Fourier
mode of the regularized scalar field, observed at a fixed ``watchpoint''
spatial position in the strong-field region.  We consider the model
\begin{widetext}
\begin{equation}
\bigl. (\varphi_m) \bigr. _\regularized
	= B(t)
	  + \sum_k A^{(k)}
		   \exp \left( - \lambda^{(k)} (t - t_\rref) \right)
		   \,
		   \sin \left(
			2\pi \frac{t - t_\rref}{P^{(k)}}
			+ \eta^{(k)}
			\right)
								\,\, \text{,}
			     \label{eqn:scalar-field=bg+sum-of-damped-sinusoids}
\end{equation}
\end{widetext}
where
$B$ is a spline function representing the slowly-varying ``background''
variation of the scalar field,
$k$~indexes the damped-sinusoids included in the model,
$A^{(k)}$, $\lambda^{(k)}$, $P^{(k)}$, and $\eta^{(k)}$ are
respectively the amplitude, damping rate, period, and phase offset
of each damped-sinusoid,
and
the subscript ${}_\rref$~denotes a ``reference'' time chosen for convenience.
To avoid degeneracy between the spline and the damped-sinusoid we require
that the spacing in $t$ between adjacent spline control points always be
at least $1.5 P^{(\max)}$, where $P^{(\max)} := \displaystyle\max_k P^{(k)}$
is the period of the longest-period damped-sinusoid in the model.


\subsubsection{Perturbations at the particle position}
\label{sect:QNM-models-and-fits/scalar-field/self-force}

Consider next the case of wiggles in the radiation-reaction self-force
(which is implicitly defined at the particle position).  This introduces
two complications: the self-force is a 4-vector (with nontrivial $t$, $r$,
and $\phi$ components for our equatorial orbits), and the field perturbation
is being sampled at a time-varying position.
Analogously to~\eqref{eqn:scalar-field=bg+sum-of-damped-sinusoids},
we thus consider the model
\begin{widetext}
\begin{equation}
F_a(u) = \frac{B_a(u)}{r^3_\particle(u)}
	 + \sum_k \frac{A^{(k)}_a}{r_\particle(u)}
		  \exp \left( - \lambda^{(k)} (u - u_\rref) \right)
		  \sin \left(
		       2\pi \frac{u - u_\rref}{P^{(k)}}
		       - m \bigl( \phi_\particle(u) - \phi_\rref \bigr)
		       + \eta^{(k)}_a
		       \right)
								\,\, \text{,}
			       \label{eqn:scalar-F-a=bg+sum-of-damped-sinusoids}
\end{equation}
\end{widetext}
where
we now parameterize the particle's motion using the retarded time
coordinate~$u$,
\footnote{
	 Heuristically, the choice of~$u$ rather than~$t$ as an
	 independent variable in the model is suggested by the
	 QNM signals propagating outward along light cones after
	 being excited near periapsis.  However, it is not clear
	 that this is a good approximation to actual QNM dynamics,
	 so we experimented with models using both~$t$ and~$u$
	 as independent variables.  We found the latter to give
	 better fits to our numerical data.
	 }
{}
$B_a$ is now a 4-vector spline function representing the background
variation of the self-force along the particle worldline,
$k$~again indexes the damped-sinusoids included in the model,
$A_a^{(k)}$, $\lambda^{(k)}$, $P^{(k)}$, and $\eta_a^{(k)}$ are now
respectively the 4-vector amplitude, damping rate, period,
and 4-vector phase offset of each damped-sinusoid,
and
the subscript ${}_\rref$~again denotes a ``reference'' time
chosen for convenience.
The non-degeneracy condition on the background spline now applies to the
spacing in~$u$ between adjacent spline control points.

Notice that the damping rate and oscillation period of each
damped-sinusoid are common across all tensor components of the self-force.
The $- m \phi_\particle(u)$~term models the variation in oscillation
phase due to particle's (i.e., the sampling point's) motion in~$\phi$.
The $1/r^3_\particle(u)$ and $1/r_\particle(u)$ factors model the
expected far-field variation in self-force and in the oscillation
eigenfunction amplitude with position.  (Actual QNM eigenfunctions
have a much more complicated variation of amplitude with spatial position,
but for simplicity we omit this from our model.)


\subsubsection{Perturbations at $\Scri^+$}
\label{sect:QNM-models-and-fits/scalar-field/Scri}

Finally, consider the case of wiggles in the radiated (physical) field
at $\Scri^+$.  Because of the hyperboloidal time slices, we observe
these at finite coordinate time~$t$ (the $\Scri^+$ time has an arbitrary
offset relative to the strong-field coordinate time).  As noted in
section~\ref{sect:field-diagnostics}, here it is somewhat difficult
to separate wiggles from the overall radiation pattern, so we consider
second time derivatives of the physical scalar-field modes.
Analogously to~\eqref{eqn:scalar-field=bg+sum-of-damped-sinusoids}
and~\eqref{eqn:scalar-F-a=bg+sum-of-damped-sinusoids}, we thus consider
the model
\begin{widetext}
\begin{equation}
\biggl.
\partial_{tt} \left( \bigl. (\varphi_m) \bigr. _\regularized \right)
\biggr|_{\Scri^+}
	= B(t)
	  + \sum_k A^{(k)}
		   \exp \left( - \lambda^{(k)} (t - t_\rref) \right)
		   \,
		   \sin \left(
			2\pi \frac{t - t_\rref}{P^{(k)}}
			+ \eta^{(k)}
			\right)
			\label{eqn:scalar-field@Scri=bg+sum-of-damped-sinusoids}
								\,\, \text{,}
\end{equation}
\end{widetext}
where the meanings of all terms
(and the non-degeneracy condition on the background spline)
are the same as in~\eqref{eqn:scalar-field=bg+sum-of-damped-sinusoids}.


\subsubsection{Fitting the models}
\label{sect:QNM-models-and-fits/scalar-field/fits}

For each of the models~\eqref{eqn:scalar-field=bg+sum-of-damped-sinusoids},
\eqref{eqn:scalar-F-a=bg+sum-of-damped-sinusoids},
and~\eqref{eqn:scalar-field@Scri=bg+sum-of-damped-sinusoids},
we visually inspect plots of our time-series data to identify a suitable
range of the independent variable for fitting and to choose initial guesses
for the background and wiggle parameters, then use the nonlinear
least-squares fitting subroutine \Subroutine{LMDIF1} from the
\SubroutineLibrary{MINPACK} library~\cite{MINPACK}
to fit the model to the data.  To make the model closer to linear
(which improves the convergence of the nonlinear fitting), we fit
the wiggle amplitudes and phases as cosine- and sine-component amplitudes
\big(i.e., $A \sin(X + \eta)$ is actually fitted as
$A^{(\cos)} \cos(X) + A^{(\sin)} \sin(X)$\big).
In most cases we used uniform weighting for the fits,
but in a few cases we used weights proportional to~$r^3$
so as to improve the fit at late times (close to apoapsis).
 

\subsubsection{Uncertainties in the fitted wiggle parameters}
\label{sect:QNM-models-and-fits/scalar-field/Monte-Carlo}

The residuals from our wiggle fits are \emph{not} random, but rather
are dominated by low-amplitude oscillations of similar frequency to
the wiggles themselves (this can be seen in
figures~\ref{fig:sf-a99p3e8-m1-fits-and-residuals}
and~\ref{fig:sf-a99p3e8-m4-fits-and-residuals}).
This means that formal uncertainty estimates for the fitted parameters
$\bigl( P^{(k)}, \tau^{(k)} \bigr)$ (derived assuming uncorrelated
Gaussian residuals) are not realistic.
Because of the oscillatory nature of the residuals, the fitted parameters
are slightly dependent on the precise choice of fitting interval;
this is, in fact, usually the dominant uncertainty in the fitted parameters.

We use a Monte-Carlo procedure to estimate realistic uncertainties
in the fitted parameters:  Given a fit of one of the above wiggle
models to our data in some interval~$I$ (in either~$t$ or~$u$) of
length~$L_\fit \ge 4 P^{(\max)}$, we randomly choose $N_{\text{trial}} = 300$
subintervals of~$I$ (randomly sampling each lower and upper interval
endpoint from a uniform distribution) subject to the constraint that
each subinterval must have a minimum length of
$L_{\min} \,{=}\, 3 P^{(\max)}$.
\footnote{
\label{footnote:sf-fits-minimum-length-requirements}
	 The minimum-length requirement for the subintervals
	 ensures that each subinterval is long enough to allow
	 a reasonable estimate of the wiggle decay rate and
	 period (the fitting errors should scale roughly
	 inversely with $L_{\min}$).
	 The minimum-length requirement for the full
	 fitting interval ensures that different subintervals
	 can sample significantly different regions of the data.
	 }
{}  Then we repeat the wiggle-model fit for each subinterval.
\footnote{
	 Each subinterval fit uses a subset of the
	 original fit's background spline control points
	 which just span that subinterval, plus one point
	 outside the subinterval on each of the interval's
	 left and right endpoints.
	 }
{}  The ensemble of the $N_{\text{trial}}$ sets of Monte-Carlo--trial
fitted parameters then provides an estimate of the uncertainty
in the fitted parameters from the full-interval fit.

After allowing initial transients to decay, our numerical calculations
extend over a number of particle orbits.  Because the particle orbit
precesses strongly, each orbit places the particle in a different
position with respect to any fixed watchpoint or $\Scri^+$~observer.
For each orbit we repeat the entire fitting procedure (including the
full set of Monte-Carlo sub-interval trials).  Our final estimate for
the uncertainty in the fitted parameters is obtained from the union
of all the Monte-Carlo trials over several (typically~3) distinct orbits.

This procedure has two main limitations:
\begin{itemize}
\item	The procedure is not applicable to cases where the
	overall fitting interval is too short (length~$L_\fit < 4P^{(\max)}$).
	(Footnote~\ref{footnote:sf-fits-minimum-length-requirements}
	outlines the reasons for this.)
\item	If a wiggle is rapidly damped, then the wiggle amplitude
	becomes very small at the right (large~$t$ or~$u$) end
	of a long fitting interval,
	so a subinterval of near-minimum length ($3 P^{(\max)}$)
	which is close to the right end of the overall fitting interval
	will have a poorly-constrained fit, yielding a large scatter
	in the fitted parameters.
\end{itemize}
These limitations are most severe when the wiggles have low amplitude
and are rapidly damped, as is the case for low Kerr spins.


\subsubsection{Other error sources}
\label{sect:QNM-models-and-fits/scalar-field/other-error-sources}

There are a number of other error sources not taken into account in
our Monte-Carlo error estimates:
\begin{itemize}
\item	Our (TW's) numerical code only computes the diagnostics
	to finite accuracy.  Comparing diagnostics between calculations
	done with different numerical resolutions, we have generally
	excluded any data where the diagnostic computed at our highest
	resolution (that shown in table~\ref{tab:sf-configurations})
	differs from that computed at the next-lower resolution
	listed in table~\ref{tab:sf-grids}
	by more than a few percent.
\item	Wiggles are not perfectly separable from the background
	variation of the diagnostics.  That is, the actual frequency
	spectra of the diagnostics are almost certainly continuous,
	and cannot be unambiguously separated into low-frequency (background)
	and high-frequency (wiggle) components.
\item	Our models for the background variation are imperfect.
	Our constraint that background spline control points
	must be spaced at least $1.5$~wiggle periods apart
	keeps the background and wiggles from being degenerate,
	but at the cost of leaving the background model
	unable to accurately fit some non-wiggle variations,
	particularly for small-$m$ (longer-period) wiggles
	where the spline control points are forced to be
	quite far apart.
\item	For wiggles in $F_a$, our wiggle
	model~\eqref{eqn:scalar-F-a=bg+sum-of-damped-sinusoids}
	doesn't accurately include the actual spatial variation
	of the wiggle (QNM) eigenfunctions.
\item	There may be multiple wiggle modes present simultaneously
	in the diagnostics for a single~$m$.  Although our wiggle
	models and fitting software support simultaneously fitting
	an arbitrary number of wiggles, we have generally not
	done this, i.e., we have generally only attempted to fit
	a single-wiggle model for each diagnostic time series.
\footnote{
	 In a few cases where our best-fitting single-wiggle
	 model's residuals showed strong systematics, we then
	 proceeded to fit 2-wiggle models.  These improved the
	 residuals by at least an order of magnitude.
	 However, all the results presented in
	 section~\ref{sect:data-and-QNM-fits/scalar-field}
	 are based on single-wiggle fits.
	 }
\end{itemize}

We believe that all these other error sources are small, but it is
difficult to quantify them.

For each wiggle fit we visually assess the fit residuals to look for
obvious systematics.  For all results reported here the fit residuals
are at least a factor of~$10$ smaller than the wiggle amplitude; in
most cases they are a factor of~$30$ to~$100$ smaller.  This suggests
that our fits are indeed accurately modelling at least the dominant
wiggle features of the diagnostics.


\subsection{Gravitational perturbations}
\label{sect:QNM-models-and-fits/gravitational}

In the gravitational case we search for the QNM frequencies in the
individual (spheroidal) $(\ssl m)$-modes. By looking at individual
$(\ssl m)$-modes we minimize the number of QNMs that need to be modelled
(fitted) simultaneously.  Since QNMs appear naturally as spheroidal modes,
using spheroidal modes minimizes the amount of ``crosstalk'' mixed in from
neighboring modes. Note that since the spheroidicity of the spheroidal
harmonics in the Teukolsky equation depends on the frequency of the modes,
the QNMs have complex spheroidicity and will not project perfectly on the
corresponding (real) spheroidal modes that appear
in the field solutions. 
Consequently, ``crosstalk'' between the modes cannot be fully
eliminated. Nonetheless, the crosstalk in the spheroidal modes should be
significantly smaller than if one were to use the spherical $\ell$-modes.


\subsubsection{Fit models}

The fit models used in the gravitational case are very similar to the ones
used in the scalar case. For the ``watchpoint'' diagnostics we use
\begin{widetext}
\begin{equation}
 \Phi_{-2}^{(\ssl m)} =\sum_n B^{(n)} t^n + \sum_k\left\{ 
 		A_s^{(k)} \sin \sqb{ \omega_k (t-t_\rref)}
 		+
		A_c^{(k)} \cos \sqb{ \omega_k (t-t_\rref)} 
	\right\}
e^{-\alpha_k (t-t_\rref)}.
\end{equation}
In this case the smooth background of the signal is modelled by a simple
polynomial in $t$. We maximize the number of linear fit parameters by
writing the model as a sum of sines and cosines.

Similarly, the $(\ssl m)$-modes contributing to the local gravitational
self-force are modelled by
\begin{equation}
F_a^{(\ssl m)} =\sum_n B^{(n)}_a u^n + \sum_k\left\{ 
	A_{s,a}^{(k)} \sin \sqb{ \omega_k (u-u_\rref)-m\phi_\particle}
	+
	A_{c,a}^{(k)} \cos \sqb{ \omega_k (u-u_\rref)-m\phi_\particle} 
\right\}\frac{e^{-\alpha_k (u-u_\rref)}}{r_\particle}.
\end{equation}
As in the scalar case, the main shortcoming in this model is the
inaccurate modelling of the QNMs' radial profiles.

Finally, the model for the gravitational waveform at~$\Scri^+$ is
very similar to the model for the watchpoints,
\begin{equation}\label{eq:gravscriplusmodel}
\lim_{r\to\infty} r\Psi_{4}^{(\ssl m)} =\sum_n B^{(n)} u^n + \sum_k\left\{ 
	A_s^{(k)} \sin \sqb{ \omega_k (u-u_\rref)}
	+
	A_c^{(k)} \cos \sqb{ \omega_k (u-u_\rref)} 
\right\}e^{-\alpha_k (u-u_\rref)}
								\,\, \text{.}
\end{equation}
\end{widetext}


\subsubsection{Fitting procedure}

The only non-linear parameters in the above models are the QNM
frequencies~$\omega_k$ and decay constants~$\alpha_k$. Consequently,
for fixed~$\omega_k$ and~$\alpha_k$ the remaining parameters can
be determined efficiently through a linear least squares procedure.
This is implemented by using \mathematica's \Subroutine{LinearModelFit}
routine for each diagnostic on a suitable time window of data. To
reduce the impact of un-modelled higher order QNMs, these fits are
weighted by $\exp(2\alpha_1 t)$. Typically, the fits include around
20~terms in the background model and up to 8~QNMs.

The $\omega_k$ and $\alpha_k$ are then determined by maximizing the
sum of the adjusted $R^2$ values of all the component fits. This
is implemented using \mathematica's \Subroutine{FindMaximum} with
the \Subroutine{PrincipleAxis} method. The initial values for
$\omega_k$ and $\alpha_k$ are set by the numerically known corresponding
QNMs offset by a random~$\O(1\%)$ amount. An indication of the
modelling error is obtained by varying the fit window and number
of background terms, and determining the spread of the best fits.


\section{Data and quasinormal-mode fits}
\label{sect:data-and-QNM-fits}


\subsection{Scalar field}
\label{sect:data-and-QNM-fits/scalar-field}

\begin{figure}[b]
\begin{center}
\includegraphics[width=\columnwidth]{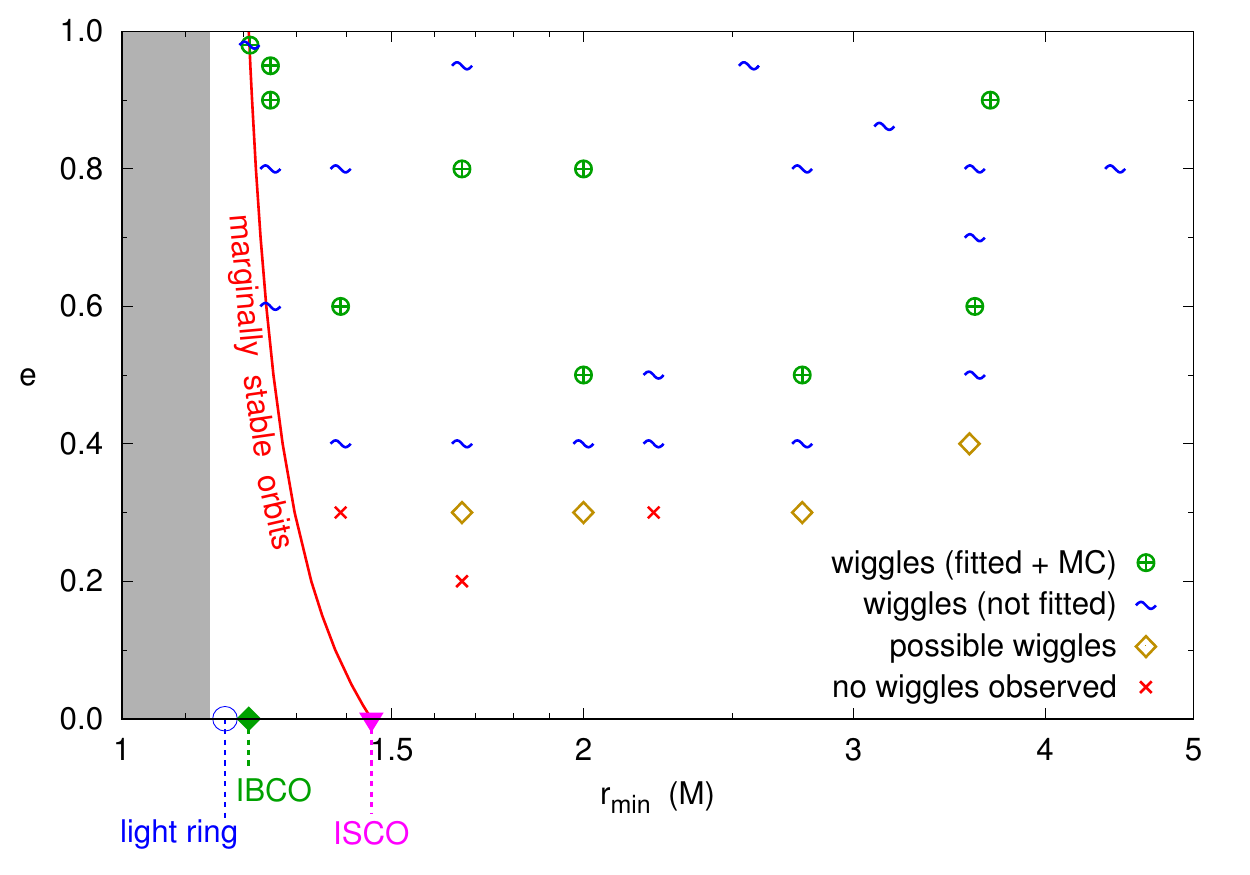}
\end{center}
\caption{
\label{fig:sf-a99-phase-space-rmin-e}
	This figure shows the phase space of the Kerr spin~$\tilde{a} = 0.99$
	scalar-field configurations presented here, plotted in terms
	of the periapsis radius $r_{\min}$ and the eccentricity~$e$.
	The shaded region at the left shows orbits with periapsis
	inside the horizon.  The light ring,
	innermost bound circular orbit (IBCO),
	innermost stable circular orbit (ISCO),
	and the locus of marginally stable orbits are also shown.
	The meanings of the plot symbols are described in detail
	in Table~\protect\ref{tab:sf-wiggle-symbols-key}.
	}
\end{figure}

We have surveyed a large number of configurations for
Kerr spin~$\tilde{a} = 0.99$, together with a smaller number of
configurations for other Kerr spins; for selected configurations
we have fitted (or attempted to fit) wiggle models as described in
section~\ref{sect:QNM-models-and-fits/scalar-field}.
Tables~\ref{tab:sf-configurations} and \ref{tab:sf-wiggle-symbols-key}
describe all the configurations surveyed here, and
figure~\ref{fig:sf-a99-phase-space-rmin-e} shows the
$(\text{periapsis radius}, \text{orbital eccentricity})$
phase space of the $\tilde{a} = 0.99$ configurations.

\begingroup
\squeezetable
\begin{table*}[p]
\begin{center}
\begin{ruledtabular}
\begin{tabular}
  {l@{\hspace{-1.0em}}d@{\hspace{-2.0em}}d@{\hspace{-2.0em}}d
%
  d@{\hspace{-1.0em}}dcdr
  c@{}c@{}cc@{}c@{}cc@{}c@{}cc@{}c@{}cc@{}c@{}cc@{}c@{}c}
Name		& \multicolumn{1}{c}{\hspace{2.5em}$\tilde{a}$}
				& \multicolumn{1}{c}{\hspace{2.5em}$p$}
						& \multicolumn{1}{c}{$e$}
		& \multicolumn{1}{c}{$T_r$}
				& \multicolumn{1}{c}{\hspace{1.25em}$r_{\min}$}
						& $\dot{\phi}_{\periapsis}$
		& \multicolumn{1}{c}{$r_\watchpoint$}
				& \multicolumn{1}{c}{Resolution}
	& \multicolumn{3}{c}{m=1}
				& \multicolumn{3}{c}{m=2}
							& \multicolumn{3}{c}{m=3}
	& \multicolumn{3}{c}{m=4}
				& \multicolumn{3}{c}{m=5}
							& \multicolumn{3}{c}{m=6}
									\\
\cline{10-12}\cline{13-15}\cline{16-18}\cline{19-21}\cline{22-24}\cline{25-27}
		&		&		&
		& \multicolumn{1}{c}{($M$)}
				& \multicolumn{1}{c}{\hspace{1.25em}($M$)}
						& ($M^{-1}$)
		& \multicolumn{1}{c}{($M$)}
				&
	& w & $F$ & $\Scri$	& w & $F$ & $\Scri$	& w & $F$ & $\Scri$
	& w & $F$ & $\Scri$	& w & $F$ & $\Scri$	& w & $F$ & $\Scri$
									\\
\hline 
w9x5-161	& 0.999\,99	& 2.918\,315	& 0.807\,519
		& 230.442	& 1.615		& 0.344\,414
		& 11.810	& dro12-96
	& \WigglesLikely	& \WigglesLikely	&
	& \WigglesLikely	& \WigglesLikely	&
	& \WigglesLikely	& \WigglesLikely	&
	& \WigglesLikely	& \WigglesLikely	&
	& \WigglesLikely	& \WigglesLikely	&
	& \WigglesLikely	& \WigglesLikely	&
									\\
\hline 
w9x4-368	& 0.999\,9	& 7.0		& 0.9
		& 1513.112	& 3.684		& 0.145\,444
		& 46.304	& dro8-64
	& \WigglesYesMC		& \WigglesYesMC		& \WigglesYesMC
	& \WigglesYesMC		& \WigglesYesMC		& \WigglesYesMC
	&			& \WigglesYesMC		& \WigglesYesMC
	&			& \WigglesYesMC		& \WigglesNo
	&			& \WigglesNo		& \WigglesNo
	&			& \WigglesNo		& \WigglesNo
									\\
\hline 
w999-278	& 0.999		& 5.0		& 0.8
		& 400.508	& 2.778		& 0.199\,135
		& 21.448	& dro10-80
	& \WigglesYesMC		& \WigglesYesMC		& \WigglesYesMC
	& \WigglesYesMC		& \WigglesYesMC		& \WigglesYesMC
	& \WigglesYesMC		& \WigglesYesMC		& \WigglesYesMC
	&			&			&
	&			&			&
	&			&			&
									\\
w999-368	& 0.999		& 7.0		& 0.9
		& 1513.179	& 3.684		& 0.145\,443
		&		& dro8-64
	&			& \WigglesYesMC		& \WigglesYesMC
	&			& \WigglesYesMC		& \WigglesYesMC
	&			& \WigglesYesMC		&
	&			& \WigglesLikely	& \WigglesNo
	&			& \WigglesNo		& \WigglesNo
	&			&			& \WigglesNo
									\\
\hline 
ze98a		& 0.99		& 2.398\,1	& 0.98
		& 3414.259	& 1.211		& 0.430\,498
		& 61.223	& dro6-48
	& \WigglesLikely	& \WigglesLikely	&
	& \WigglesLikely	& \WigglesLikely	&
	& \WigglesLikely	& \WigglesLikely	&
	& \WigglesLikely	& \WigglesLikely	&
	& \WigglesLikely	& \WigglesLikely	&
	& \WigglesLikely	& \WigglesLikely	&
									\\
ze98		& 0.99		& 2.4		& 0.98
		& 3304.620	& 1.212		& 0.430\,300
		& 		& dro10-80
	&			& \WigglesYesMC		& \WigglesYesMC
	&			& \WigglesYesMC		& \WigglesYesMC
	&			& \WigglesYesMC		& \WigglesYesMC
	&			& \WigglesYesMC		& \WigglesYesMC
	&			& \WigglesYesMC		& \WigglesYesMC
	&			& \WigglesYesMC		& \WigglesLikely
									\\
w99-125a	& 0.99		& 2.437\,5  	& 0.95
		& 957.757	& 1.25		& 0.421\,924
		& 36.314	& dro8-64
	& \WigglesYesMC		& \WigglesYesMC		& \WigglesYesMC
	& \WigglesYesMC		& \WigglesYesMC		& \WigglesYesMC
	& \WigglesYesMC		& \WigglesYesMC		& \WigglesYesMC
	&			& \WigglesLikely	& \WigglesLikely
	&			&			&
	&			&			&
									\\
w99-125b	& 0.99		& 2.375		& 0.9
		& 432.084	& 1.25		& 0.421\,555
		& 20.543	& dro8-64
	& \WigglesYesMC		& \WigglesYesMC		& \WigglesYesMC
	& \WigglesYesMC		& \WigglesYesMC		& \WigglesYesMC
	& \WigglesYesMC		& \WigglesYesMC		& \WigglesYesMC
	& \WigglesYesMC		& \WigglesYesMC		& \WigglesYesMC
	&			& \WigglesLikely	&
	&			&			&
									\\
w99-125c	& 0.99		& 2.25		& 0.8
		& 247.845	& 1.25		& 0.420\,811
		& 8.595		& dro6-48
	& \WigglesLikely	& \WigglesLikely	&
	& \WigglesLikely	& \WigglesLikely	&
	& \WigglesLikely	& \WigglesLikely	&
	& \WigglesLikely	& \WigglesLikely	&
	& \WigglesLikely	& \WigglesLikely	&
	& \WigglesLikely	& \WigglesLikely	&
									\\
w99-125d	& 0.99		& 2.0		& 0.6
		& 228.354	& 1.25		& 0.419\,300
		& 3.003		& dro8-64
	& \WigglesLikely	& \WigglesLikely	&
	& \WigglesLikely	& \WigglesLikely	&
	& \WigglesLikely	& \WigglesLikely	&
	& \WigglesLikely	& \WigglesLikely	&
	& \WigglesLikely	& \WigglesLikely	&
	&			&			&
									\\
w99-139		& 0.99		& 2.5		& 0.8
		& 211.605	& 1.389		& 0.389\,231
		&		& dro8-64
	&			& \WigglesLikely	&
	&			& \WigglesLikely	&
	&			& \WigglesLikely	&
	&			& \WigglesLikely	&
	&			& \WigglesLikely	&
	&			& \WigglesLikely	&
									\\
w99-139b	& 0.99		& 2.222\,222	& 0.6
		& 133.939	& 1.389		& 0.386\,760
		& 3.926		& dro8-64
	& \WigglesYes		& \WigglesYes		& \WigglesYes
	& \WigglesYes		& \WigglesYesMC		& \WigglesYesMC
	& \WigglesYesMC		& \WigglesYesMC		& \WigglesYesMC
	& \WigglesYesMC		& \WigglesYesMC		& \WigglesYesMC
	& \WigglesLikely	&			& \WigglesLikely
	&			&			&
									\\
w99-139d	& 0.99		& 1.944\,444	& 0.4
		& 126.071	& 1.389		& 0.384\,220
		& 1.984		& dro6-48
	& \WigglesNo		& \WigglesPossible	&
	& \WigglesNo		& \WigglesLikely	&
	& \WigglesNo		& \WigglesLikely	&
	& \WigglesNo		&			&
	& \WigglesNo		&			&
	& \WigglesNo		&			&
									\\
w99-139c	& 0.99		& 1.805\,556	& 0.3
		& 135.092	& 1.389		& 0.382\,960
		& 1.743		& dro6-48
	& \WigglesNo		& \WigglesNo		&
	& \WigglesNo		& \WigglesNo		&
	& \WigglesNo		& \WigglesNo		&
	& \WigglesNo		& \WigglesNo		&
	& \WigglesNo		& \WigglesNo		&
	& \WigglesNo		& \WigglesNo		&
									\\
w99-167m	& 0.99		& 3.25		& 0.95
		& 1329.680	& 1.667		& 0.336\,304
		& 47.740	& dro8-64
	& \WigglesLikely	& \WigglesLikely	&
	& \WigglesLikely	& \WigglesLikely	&
	& \WigglesLikely	& \WigglesLikely	&
	& \WigglesLikely	& \WigglesLikely	&
	& \WigglesLikely	& \WigglesLikely	&
	&			&			&
									\\
w99-167		& 0.99		& 3.0		& 0.8
		& 230.442	& 1.667		& 0.333\,682
		& 11.810	& dro10-80
	& \WigglesYesMC		& \WigglesYesMC		& \WigglesYesMC
	& \WigglesYesMC		& \WigglesYesMC		& \WigglesYesMC
	& \WigglesYesMC		& \WigglesYesMC		& \WigglesYesMC
	& \WigglesYesMC		& \WigglesYesMC		& \WigglesYesMC
	& \WigglesYesMC		& \WigglesYesMC		& \WigglesYesMC
	& \WigglesYesMC		& \WigglesYesMC		& \WigglesYesMC
									\\
w99-167k	& 0.99		& 2.333\,333	& 0.4
		& 100.014	& 1.667		& 0.326\,143
		& 2.332		& dro6-48
	& \WigglesPossible	& \WigglesPossible	&
	& \WigglesNo		& \WigglesLikely	&
	& \WigglesNo		& \WigglesLikely	&
	& \WigglesNo		& \WigglesPossible	&
	& \WigglesNo		& \WigglesNo		&
	& \WigglesNo		&			&
									\\
w99-167d	& 0.99		& 2.166\,667	& 0.3
		& 97.092	& 1.667		& 0.324\,141
		& 2.332		& dro6-48
	& \WigglesNo		& \WigglesPossible	&
	& \WigglesNo		& \WigglesPossible	&
	& \WigglesNo		& \WigglesPossible	&
	& \WigglesNo		& \WigglesNo		&
	& \WigglesNo		& \WigglesNo		&
	& \WigglesNo		& \WigglesNo		&
									\\
w99-167j	& 0.99		& 2		& 0.2
		& 98.818	& 1.667		& 0.322\,115
		& 1.984		& dro6-48
	& \WigglesNo		& \WigglesNo		&
	& \WigglesNo		& \WigglesNo		&
	& \WigglesNo		& \WigglesNo		&
	& \WigglesNo		& \WigglesNo		&
	& \WigglesNo		& \WigglesNo		&
	& \WigglesNo		& \WigglesNo		&
									\\
w99-200d	& 0.99		& 3.6		& 0.8
		& 273.551	& 2.0		& 0.281\,145
		& 14.784	& dro8-64
	& \WigglesYesMC		& \WigglesYesMC		& \WigglesYesMC
	& \WigglesYesMC		& \WigglesYesMC		& \WigglesYesMC
	& \WigglesYesMC		& \WigglesYesMC		& \WigglesYesMC
	& \WigglesYesMC		& \WigglesYesMC		& \WigglesYesMC
	& \WigglesYesMC		&			&
	&			&			&
									\\
w99-200		& 0.99		& 3.0		& 0.5
		& 111.575	& 2.000		& 0.274\,441
		&		& dro6-48
	&			& \WigglesYesMC		&
	&			& \WigglesYesMC		&
	&			& \WigglesYesMC		&
	&			& \WigglesYesMC		&
	&			& \WigglesNo		&
	&			& \WigglesNo		&
									\\
w99-200b	& 0.99		& 2.8		& 0.4
		& 99.000	& 2.000		& 0.272\,062
		& 3.003		& dro6-48
	& \WigglesNo		& \WigglesLikely	&
	& \WigglesNo		& \WigglesLikely	&
	& \WigglesNo		& \WigglesLikely	&
	& \WigglesNo		& \WigglesPossible	&
	& \WigglesNo		& \WigglesNo		&
	& \WigglesNo		& \WigglesNo		&
									\\
w99-200c	& 0.99		& 2.6		& 0.3
		& 91.817	& 2.000		& 0.269\,607
		& 2.636		& dro6-48
	& \WigglesNo		& \WigglesPossible	&
	& \WigglesNo		& \WigglesNo		&
	& \WigglesNo		& \WigglesNo		&
	& \WigglesNo		& \WigglesNo		&
	& \WigglesNo		& \WigglesNo		&
	& \WigglesNo		& \WigglesNo		&
									\\
w99-222b	& 0.99		& 3.333\,333	& 0.5
		& 117.908	& 2.222		& 0.245\,832
		& 4.193		& dro6-48
	& \WigglesLikely	& \WigglesLikely	&
	& \WigglesLikely	& \WigglesLikely	&
	& \WigglesPossible	& \WigglesLikely	&
	& \WigglesNo		&			&
	& \WigglesNo		&			&
	& \WigglesNo		&			&
									\\
w99-222		& 0.99		& 3.111\,111	& 0.4
		& 102.677	& 2.222		& 0.243\,319
		& 3.926		& dro6-48
	& \WigglesLikely	& \WigglesLikely	&
	& \WigglesNo		& \WigglesLikely	&
	& \WigglesNo		& \WigglesLikely	&
	& \WigglesNo		& \WigglesNo		&
	& \WigglesNo		& \WigglesNo		&
	& \WigglesNo		& \WigglesNo		&
									\\
w99-222c	& 0.99		& 2.888\,889	& 0.3
		& 93.395	& 2.222		& 0.240\,719
		& 3.003		& dro6-48
	& \WigglesNo		& \WigglesNo		&
	& \WigglesNo		& \WigglesNo		&
	& \WigglesNo		& \WigglesNo		&
	& \WigglesNo		& \WigglesNo		&
	& \WigglesNo		& \WigglesNo		&
	& \WigglesNo		& \WigglesNo		&
									\\
e95		& 0.99		& 5.0		& 0.95
		& 2436.050	& 2.564		& 0.220\,775
		&		& dro8-64
	&			& \WigglesLikely	&
	&			& \WigglesLikely	&
	&			& \WigglesLikely	&
	&			& \WigglesLikely	&
	&			& \WigglesLikely	&
	&			&			&
									\\
w99-278		& 0.99		& 5.0		& 0.8
		& 401.302	& 2.778		& 0.199\,076
		& 20.543	& dro8-64
	& \WigglesLikely	& \WigglesLikely	&
	& \WigglesLikely	& \WigglesLikely	&
	& \WigglesLikely	& \WigglesLikely	&
	& \WigglesLikely	& \WigglesLikely	&
	& \WigglesNo		& \WigglesLikely	&
	&			&			&
									\\
w99-278b	& 0.99		& 4.166\,667	& 0.5
		& 139.595	& 2.778		& 0.191\,854
		& 5.712		& dro8-64
	& \WigglesYesMC		& \WigglesYesMC		& \WigglesYesMC
	& \WigglesPossible	& \WigglesYesMC		& \WigglesYesMC
	& \WigglesNo		& \WigglesYesMC		& \WigglesNo
	& \WigglesNo		& \WigglesNo		& \WigglesNo
	& \WigglesNo		& \WigglesNo		& \WigglesNo
	& \WigglesNo		& \WigglesNo		& \WigglesNo
									\\
w99-278c	& 0.99		& 3.888\,889	& 0.4
		& 117.842	& 2.778		& 0.189\,278
		& 4.193		& dro6-48
	& \WigglesPossible	& \WigglesLikely	&
	& \WigglesNo		& \WigglesLikely	&
	& \WigglesNo		& \WigglesNo		&
	& \WigglesNo		& \WigglesNo		&
	& \WigglesNo		& \WigglesNo		&
	& \WigglesNo		& \WigglesNo		&
									\\
w99-278d	& 0.99		& 3.611\,111	& 0.3
		& 103.835	& 2.778		& 0.186\,607
		& 3.672		& dro6-48
	& \WigglesNo		& \WigglesPossible	&
	& \WigglesNo		& \WigglesNo		&
	& \WigglesNo		& \WigglesNo		&
	& \WigglesNo		& \WigglesNo		&
	& \WigglesNo		& \WigglesNo		&
	& \WigglesNo		& \WigglesNo		&
									\\
s99		& 0.99		& 5.850\,762	& 0.861\,866
		& 771.968	& 3.142		& 0.174\,409
		&		& dro6-48
	&			& \WigglesLikely	&
	&			& \WigglesLikely	&
	&			& \WigglesLikely	&
	&			& \WigglesLikely	&
	&			&			&
	&			&			&
									\\
w99-357		& 0.99		& 5.0		& 0.4
		& 146.751	& 3.571		& 0.139\,748
		&		& dro6-48
	&			& \WigglesPossible	&
	&			& \WigglesNo		&
	&			& \WigglesNo		&
	&			& \WigglesNo		&
	&			& \WigglesNo		&
	&			& \WigglesNo		&
									\\
w99-360c	& 0.99		& 6.48		& 0.8
		& 560.918	& 3.6		& 0.147\,411
		&		& dro6-48
	&			& \WigglesLikely	&
	&			& \WigglesLikely	&
	&			& \WigglesLikely	&
	&			& \WigglesNo		&
	&			&
	&			&
									\\
w99-360b	& 0.99		& 6.12		& 0.7
		& 330.647	& 3.6		& 0.145\,260
		&		& dro6-48
	&			& \WigglesLikely	&
	&			& \WigglesLikely	&
	&			& \WigglesLikely	&
	&			& \WigglesNo		&
	&			& \WigglesNo		&
	&			& \WigglesNo		&
									\\
w99-360a	& 0.99		& 5.76		& 0.6
		& 232.413	& 3.6		& 0.143\,040
		& 11.810	& dro8-64
	& \WigglesYesMC		& \WigglesYesMC		& \WigglesYesMC
	& \WigglesNo		& \WigglesYesMC		& \WigglesYesMC
	& \WigglesNo		& \WigglesNo		& \WigglesNo
	& \WigglesNo		& \WigglesNo		& \WigglesNo
	& \WigglesNo		& \WigglesNo		& \WigglesNo
	& \WigglesNo		& \WigglesNo		& \WigglesNo
									\\
w99-360j	& 0.99		& 5.4		& 0.5
		& 179.842	& 3.6		& 0.140\,745
		& 7.835		& dro6-48
	& \WigglesPossible	& \WigglesLikely	&
	& \WigglesNo		& \WigglesNo
	& \WigglesNo		& \WigglesNo
	& \WigglesNo		& \WigglesNo
	& \WigglesNo		&
	& \WigglesNo		&
									\\
e9		& 0.99		& 7.0		& 0.9
		& 1513.855	& 3.684		& 0.145\,429
		& 25.105	& dro8-64
	& \WigglesYesMC		& \WigglesYesMC		& \WigglesYesMC
	& \WigglesYesMC		& \WigglesYesMC		& \WigglesYesMC
	& \WigglesNo		& \WigglesYesMC		& \WigglesYesMC
	& \WigglesNo		& \WigglesYesMC		& \WigglesNo
	& \WigglesNo		& \WigglesNo		& \WigglesNo
	& \WigglesNo		& \WigglesNo		& \WigglesNo
									\\
w99-444		& 0.99		& 8.0		& 0.8
		& 745.170	& 4.444		& 0.113\,763
		&		& dro6-48
	&			& \WigglesLikely	&
	&			& \WigglesNo		&
	&			& \WigglesNo		&
	&			& \WigglesNo		&
	&			& \WigglesNo		&
	&			& \WigglesNo		&
									\\
\hline 
w95-368		& 0.95		& 7.0		& 0.9
		& 1516.962	& 3.684		& 0.145\,349
		&		& dro6-48
	&			& \WigglesYesMC		&
	&			& \WigglesYesMC		&
	&			& \WigglesYesMC		&
	&			& \WigglesYesMC		&
	&			& \WigglesNo		&
	&			& \WigglesNo		&
									\\
\hline 
n96		& 0.9		& 5.5		& 0.6
		& 227.038	& 3.437\,5	& 0.151\,199
		&		& dro8-64
	&			& \WigglesLikely		&
	&			& \WigglesLikely		&
	&			& \WigglesNo		&
	&			& \WigglesNo		&
	&			& \WigglesNo		&
	&			& \WigglesNo		&
									\\
w9-368		& 0.9		& 7.0		& 0.9
		& 1521.097	& 3.684		& 0.145\,202
		&		& dro6-48
	&			& \WigglesYesMC		& \WigglesYesMC
	&			& \WigglesYesMC		& \WigglesYesMC
	&			& \WigglesYesMC		& \WigglesYesMC
	&			& \WigglesYesMC		& \WigglesYesMC
	&			& \WigglesNo		& \WigglesNo
	&			& \WigglesNo		& \WigglesNo
									\\
n95		& 0.9		& 10.0		& 0.5
		& 378.408	& 6.667		& 0.062\,994
		&		& dro8-64
	&			& \WigglesNo		&
	&			& \WigglesNo		&
	&			& \WigglesNo		&
	&			& \WigglesNo		&
	&			& \WigglesNo		&
	&			& \WigglesNo		&
									\\
\hline 
w8-368		& 0.8		& 7.0		& 0.9
		& 1530.314	& 3.684		& 0.144\,751
		& 18.300	& dro6-48
	& \WigglesNo		& \WigglesYesMC		& \WigglesYes
	& \WigglesNo		& \WigglesYesMC		& \WigglesYesMC
	& \WigglesNo		& \WigglesYesMC		& \WigglesYesMC
	& \WigglesNo		& \WigglesYes		& \WigglesYesMC
	& \WigglesNo		&			& \WigglesNo
	& \WigglesNo		&			& \WigglesNo
									\\
e8b		& 0.8		& 8.0		& 0.8
		& 756.641	& 4.444		& 0.113\,578
		&		& dro6-48
	&			& \WigglesLikely	&
	&			& \WigglesNo		&
	&			& \WigglesNo		&
	&			& \WigglesNo		&
	&			& \WigglesNo		&
	&			& \WigglesNo		&
									\\
\hline 
e8		& 0.6		& 8.0		& 0.8
		& 771.968	& 4.444		& 0.113\,000
		&		& dro8-64
	&			& \WigglesLikely		&
	&			& \WigglesNo		&
	&			& \WigglesNo		&
	&			& \WigglesNo		&
	&			& \WigglesNo		&
	&			& \WigglesNo		&
									\\
\hline 
w4-368		& 0.4		& 7.0		& 0.9
		& 1588.133	& 3.684		& 0.140\,816
		&		& dro6-48
	&			& \WigglesYes		& \WigglesYes
	& 			& \WigglesYesMC		& \WigglesYesMC
	& 			& \WigglesYesMC		& \WigglesYes
	& 			& \WigglesYesMC		& \WigglesNo
	& 			& \WigglesLikely	& \WigglesNo
	& 			& \WigglesNo		& \WigglesNo
									\\
\hline 
w2-368		& 0.2		& 7.0		& 0.9
		& 1712.163	& 3.684		& 0.137\,622
		&		& dro6-48
	&			& \WigglesYes		& \WigglesNo
	&			& \WigglesYes		& \WigglesNo
	&			& \WigglesNo		& \WigglesNo
	&			& \WigglesNo		& \WigglesNo
	&			& \WigglesNo		& \WigglesNo
	&			& \WigglesNo		& \WigglesNo
									\\
ze4		& 0.2		& 6.15		& 0.4
		& 354.628	& 4.393		& 0.106\,691
		&		& dro6-48
	&			& \WigglesNo		&
	&			& \WigglesNo		&
	&			& \WigglesNo		&
	&			& \WigglesNo		&
	&			& \WigglesNo		&
	&			& \WigglesNo		&
									\\
\hline 
zze9		& 0.0		& 7.800\,001	& 0.9
		& 2224.815	& 4.105		& 0.120\,223
		&		& dro8-64
	&			& \WigglesNo		&
	&			& \WigglesNo		&
	&			& \WigglesNo		&
	&			& \WigglesNo		&
	&			& \WigglesNo		&
	&			& \WigglesNo		&
									\\
ze9		& 0.0		& 7.800\,1	& 0.9
		& 2112.079	& 4.105		& 0.120\,222
		&		& dro6-48
	&			& \WigglesNo		&
	&			& \WigglesNo		&
	&			& \WigglesNo		&
	&			& \WigglesNo		&
	&			& \WigglesNo		&
	&			& \WigglesNo		&
									\\
ns5		& 0.0		& 7.2		& 0.5
		& 405.662	& 4.8		& 0.095\,855
		&		& dro8-64
	&			& \WigglesNo		&
	&			& \WigglesNo		&
	&			& \WigglesNo		&
	&			& \WigglesNo		&
	&			& \WigglesNo		&
	&			& \WigglesNo		&
									\\
s0		& 0.0		& 10.695\,207	& 0.708\,941
		& 771.968	& 6.258		& 0.070\,830
		&		& dro4-32
	&			& \WigglesNo		&
	&			& \WigglesNo		&
	&			& \WigglesNo		&
	&			& \WigglesNo		&
	&			& \WigglesNo		&
	&			& \WigglesNo		&
									\\
\hline 
n-55		& -0.5		& 10.0		& 0.5
		& 505.428	& 6.667		& 0.062\,012
		&		& dro6-48
	&			& \WigglesNo		&
	&			& \WigglesNo		&
	&			& \WigglesNo		&
	&			& \WigglesNo		&
	&			& \WigglesNo		&
	&			& \WigglesNo		&
									\\
\hline 
s-6		& -0.6		& 13.083\,066	& 0.609\,412 
		& 771.968	& 8.129		& 0.048\,498
		&		& dro4-32
	&			& \WigglesNo		&
	&			& \WigglesNo		&
	&			& \WigglesNo		&
	&			& \WigglesNo		&
	&			& \WigglesNo		&
	&			& \WigglesNo		&
									\\
\hline 
wm8-631		& -0.8		& 10.1		& 0.6
		& 747.545	& 6.313		& 0.066\,625
		&		& dro4-32
	&			& \WigglesNo		&
	&			& \WigglesNo		&
	&			& \WigglesNo		&
	&			& \WigglesNo		&
	&			& \WigglesNo		&
	&			& \WigglesNo		&
									\\
\hline 
wm99-605	& -0.99		& 11.5		& 0.9
		& 3401.251	& 6.053		& 0.072\,312
		&		& dro4-32
	&			& \WigglesNo		&
	&			& \WigglesNo		&
	&			& \WigglesNo		&
	&			& \WigglesNo		&
	&			& \WigglesNo		&
	&			& \WigglesNo		&
									\\
s-99		& -0.99		& 14.542\,929	& 0.534\,714
		& 771.968	& 9.476		& 0.038\,531
		&		& dro4-32
	&			& \WigglesNo		&
	&			& \WigglesNo		&
	&			& \WigglesNo		&
	&			& \WigglesNo		&
	&			& \WigglesNo		&
	&			& \WigglesNo		&
\end{tabular}
\end{ruledtabular}
\end{center}
\vspace*{-2ex}
\caption{
\label{tab:sf-configurations}
	This table describes the Kerr scalar-field configurations
	surveyed here.
	The table rows are ordered by decreasing Kerr spin~$\tilde{a}$
	(horizontal lines separate groups of rows for the same spin),
	within this by increasing particle periapsis radius $r_{\min}$,
	and within this by decreasing particle orbital eccentricity~$e$.
	$(\tilde{a}, p, e)$ describe the configuration.
	$T_r$ is the coordinate-time period of the orbit's radial motion,
	$r_{\min}$ is the particle's periapsis $r$~coordinate,
	and $\dot{\phi}_\periapsis$ is the particle's angular 3-velocity
	in $\phi$ (i.e., $d\phi/dt$) at the point of periapsis.
	$r_\watchpoint$ is the $r$~coordinate of the watchpoint
	used for observing wiggles, or blank if no watchpoint data
	was available for this configuration.
	``Resolution'' is the highest numerical resolution used for
	simulations of this configuration, and refers to the grid structures
	described in table~\protect\ref{tab:sf-grids}.  
	The final sets of columns describe the presence or absence
	of wiggles in our field diagnostics for $m=1$ through $m=6$;
	for each $m$ there are 3~columns labeled
	``w'', ``$F$'', and ``$\Scri$'',
	describing (respectively) wiggles in the regularized field
	$\bigl. (\varphi_m) \bigr. _\regularized$ at a fixed ``watchpoint''
	(located on the equator at $r = r_\watchpoint$),
	wiggles in the self-force $F_a$,
	and wiggles in the second time derivative of the physical field,
	$\partial_{tt} \left( \bigl. (\varphi_m) \bigr. _\physical \right)$,
	at $\Scri^+$.
	The meanings of the symbols in these columns are described
	in detail in table~\protect\ref{tab:sf-wiggle-symbols-key};
	briefly,
	\WigglesYesMC{} or \WigglesYes{} means that we observed
	wiggles and fit them to the appropriate model described in
	section~\protect\ref{sect:QNM-models-and-fits/scalar-field},
	\WigglesLikely{}~means that we observed wiggle-like oscillations
	but did not attempt a quantitative fit,
	\WigglesPossible{}~means that we observed oscillations which
	might be wiggles but were not clearly separated from the
	background variation, and
	\WigglesNo{}~means that we did not observe wiggles
	(\nb{} this does \emph{not} prove the absence of wiggles,
	only that we did not observe them).
	}
\end{table*}
\endgroup

\begin{table}
\begin{center}
\begin{sloppypar}
\begin{ruledtabular}
\begin{tabular}{lp{22em}}
Symbol		& Meaning
									\\
\hline 
\WigglesYesMC	& we observed oscillations in the diagnostics,
		  successfully fit the appropriate wiggle model described in
		  section~\protect\ref{sect:QNM-models-and-fits/scalar-field}
		  over a $t$ or $u$~range of $\ge 4$~wiggle periods,
		  and performed the Monte-Carlo error analysis described in
		  section~\protect\ref{sect:QNM-models-and-fits/scalar-field/Monte-Carlo}.
									\\[1ex]
\WigglesYes	& we observed oscillations in the diagnostics
		  and successfully fit the appropriate wiggle model
		  described in
		  section~\protect\ref{sect:QNM-models-and-fits/scalar-field},
		  but the model was fitted over too short a
		  $t$ or $u$~range ($< 4$~wiggle periods) for the
		  Monte-Carlo error analysis described in
		  section~\protect\ref{sect:QNM-models-and-fits/scalar-field/Monte-Carlo}
		  to be used
									\\[1ex]
\WigglesLikely	& we observed oscillations in the diagnostics
		  which visually appeared to be wiggles,
		  with physically reasonable period and decay rate,
		  but we did not attempt to quantitatively fit a wiggle model
									\\[1ex]
\WigglesPossible& we observed oscillations in the diagnostics
		  which might have been wiggles,
		  but these oscillations were not clearly separated
		  from the background variation in the diagnostics
									\\[1ex]
\WigglesNo	& we did not observe wiggle-like oscillations
		  in the diagnostics; this could mean either that
		  no wiggles are present, or alternatively that wiggles
		  were present but they were at too low an amplitude
		  and/or insufficiently separated from the background
		  variation to be observed
									\\[1ex]
(blank)		& diagnostics were not computed, not computed
		  sufficiently accurately for studying wiggles,
		  or were computed but not assessed
\end{tabular}
\end{ruledtabular}
\end{sloppypar}
\end{center}
\caption{
\label{tab:sf-wiggle-symbols-key}
	This table explains the meanings of the ``wiggle symbols''
	used in table~\protect\ref{tab:sf-configurations}
	and figure~\protect\ref{fig:sf-a99-phase-space-rmin-e}.
	}
\end{table}

\begin{table}[b]
\begin{center}
\begin{ruledtabular}
\begin{tabular}{lD{/}{/}{-1}D{/}{/}{-1}D{/}{/}{-1}D{/}{/}{-1}}
	& \multicolumn{2}{c}{coarsest grid}
	& \multicolumn{2}{c}{finest grid}
									\\
\cline{2-3}\cline{4-5}
	& \multicolumn{1}{c}{$R_*$}
			& \multicolumn{1}{c}{$\theta$}
					& \multicolumn{1}{c}{$R_*$}
							& \multicolumn{1}{c}{$\theta$}
									\\
	& \multicolumn{1}{c}{($M$)}
			& \multicolumn{1}{c}{(radians)}
					& \multicolumn{1}{c}{($M$)}
							& \multicolumn{1}{c}{(radians)}
									\\
\hline 
dro12-96& 1/12		& \pi/216	& 1/96	& \pi/1728		\\
dro10-80& 1/10		& \pi/180	& 1/80	& \pi/1440		\\
dro8-64	& 1/8		& \pi/144	& 1/64	& \pi/1152		\\
dro6-48	& 1/6		& \pi/108	& 1/48	& \pi/864		\\
dro4-32	& 1/4		& \pi/72	& 1/32	& \pi/576		
\end{tabular}
\end{ruledtabular}
\end{center}
\caption{
\label{tab:sf-grids}
	This table gives the resolution in~$r_*$ and~$\theta$
	of the coarsest and finest grids for each of our standard
	grid structures.  The finest grid covers a neighborhood
	of the worldtube in each slice, while successively coarser
	grids cover correspondingly larger regions using standard
	Berger-Oliger $2\,{:}\,1$~mesh refinement.
	}
\end{table}

\begin{figure*}[p]
\begin{center}
\includegraphics[width=\textwidth]{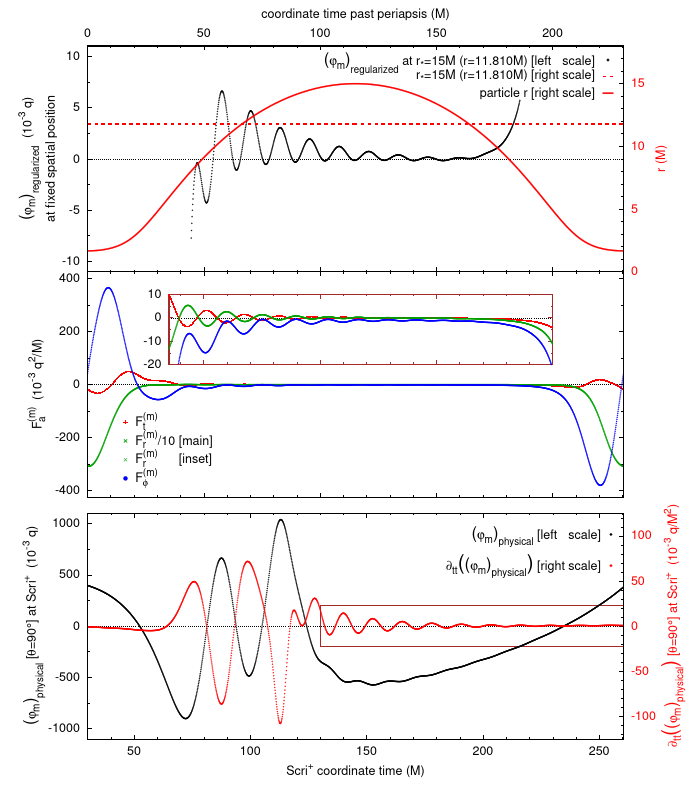}
\end{center}
\caption{
\label{fig:sf-a99p3e8-m1-wiggles-overview}
	This figure shows wiggles in the $m=1$ scalar-field diagnostics
	for a single particle orbit ($t \in [38T_r, 39T_r]$) of the
	$(\tilde{a},p,e) = (0.99,3,0.8)$ configuration.
	In the lower two subfigures the inset
	boxes (which share common time scales with their corresponding
	main figures) show the regions where wiggles are visible.
	The upper subfigure shows the regularized scalar field
	at $r_* = 15M$ ($r = 11.810M$),
	at times when this position is within the worldtube.
	The middle subfigure shows the 3~nontrivial coordinate components
	of the 4-vector self-force $F_a^{(m)}$.  The lower subfigure
	shows the physical scalar field and its second time derivative,
	on the equator at $\Scri^+$;
	the fields at other viewing angles are very similar in shape
	but with smaller amplitudes.  Due to the orbital precession,
	the $\Scri^+$ fields are \emph{not} ``periodic with period $T_r$''.
	Note that the zero point of the $\Scri^+$ time coordinate
	does not correspond to periapsis.
	}
\end{figure*}

\begin{figure*}[p]
\begin{center}
\includegraphics[width=\textwidth]{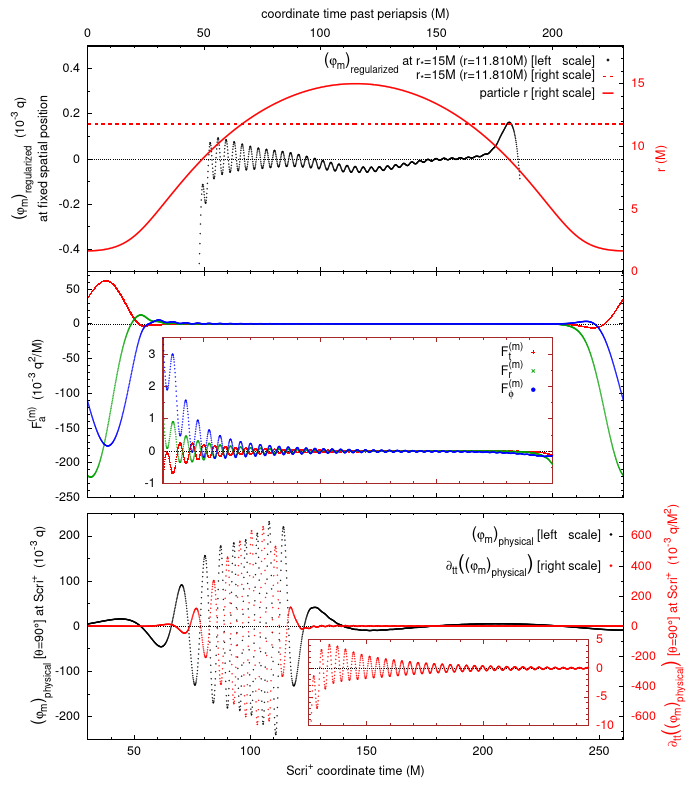}
\end{center}
\caption{
\label{fig:sf-a99p3e8-m4-wiggles-overview}
	This figure shows wiggles in the $m=4$ scalar-field diagnostics
	for a single particle orbit ($t \in [13T_r, 14T_r]$) of the
	$(\tilde{a},p,e) = (0.99,3,0.8)$ configuration.
	In the lower two subfigures the inset
	boxes (which share common time scales with their corresponding
	main figures) show the regions where wiggles are visible.
	The upper subfigure shows the regularized scalar field
	at $r_* = 15M$ ($r = 11.810M$),
	at times when this position is within the worldtube.
	The middle subfigure shows the 3~nontrivial coordinate components
	of the 4-vector self-force $F_a^{(m)}$.  The lower subfigure
	shows the physical scalar field and its second time derivative,
	on the equator at $\Scri^+$;
	the fields at other viewing angles are very similar in shape
	but with smaller amplitudes.  Due to the orbital precession,
	the $\Scri^+$ fields are \emph{not} ``periodic with period $T_r$''.
	Note that the zero point of the $\Scri^+$ time coordinate
	does not correspond to periapsis.
	}
\end{figure*}

Figures~\ref{fig:sf-a99p3e8-m1-wiggles-overview}
and~\ref{fig:sf-a99p3e8-m4-wiggles-overview} show the wiggles in the
scalar-field diagnostics for the $(\tilde{a},p,e) = (0.99,3,0.8)$ configuration,
for $m=1$ and $m=4$ respectively.  Notice that the wiggles are
visible in \emph{all} the field diagnostics.
Notice also the much higher frequency and smaller amplitude of the
$m=4$ wiggles.

\begin{figure*}[p]
\begin{center}
\includegraphics[width=\textwidth]{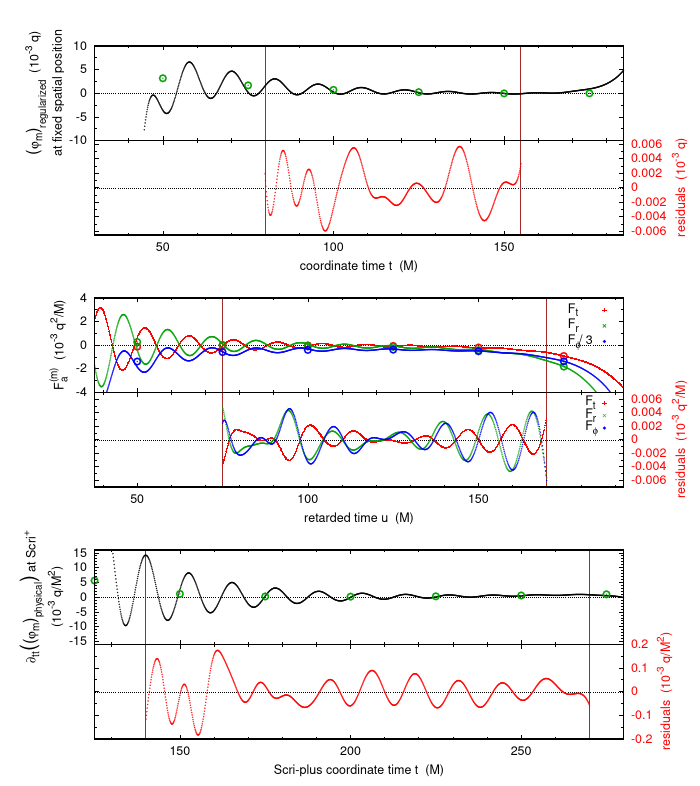}
\end{center}
\caption{
\label{fig:sf-a99p3e8-m1-fits-and-residuals}
	This figure shows the wiggles and fit residuals for the $m=1$
	scalar-field diagnostics
	for a single particle orbit ($t \in [38T_r, 39T_r]$) of the
	$(\tilde{a},p,e) = (0.99,3,0.8)$ configuration.
	In the upper subfigure, the upper plot shows the regularized
	scalar field at $r_* = 15M$
	($r = 11.810M$) as a function of coordinate time~$t$,
	at times when this position is within the worldtube,
	while the lower plot shows the residuals after fitting this data to
	the model~\protect\eqref{eqn:scalar-field=bg+sum-of-damped-sinusoids}.
	In the middle subfigure, the upper plot shows the 3~nontrivial
	coordinate components of the 4-vector self-force $F_a^{(m)}$
	as a function of retarded time $u$,
	while the lower plot shows the residuals after fitting this data to
	the model~\protect\eqref{eqn:scalar-F-a=bg+sum-of-damped-sinusoids}.
	In the lower subfigure, the upper plot shows the second time
	derivative of the physical scalar field on the equator at $\Scri^+$
	as a function of the $\Scri^+$~coordinate time~$t$,
	while the lower plot shows the residuals after fitting this data to the
	model~\protect\eqref{eqn:scalar-field@Scri=bg+sum-of-damped-sinusoids}.
	In each subfigure the vertical tan lines mark the interval
	over which the model is fitted, the circles in the upper subfigure
	show the background spline control points, and the differences
	between the data and the fitted model are too small to be visible
	on the scale of the upper plot.
	}
\end{figure*}

\begin{figure*}[p]
\begin{center}
\includegraphics[width=\textwidth]{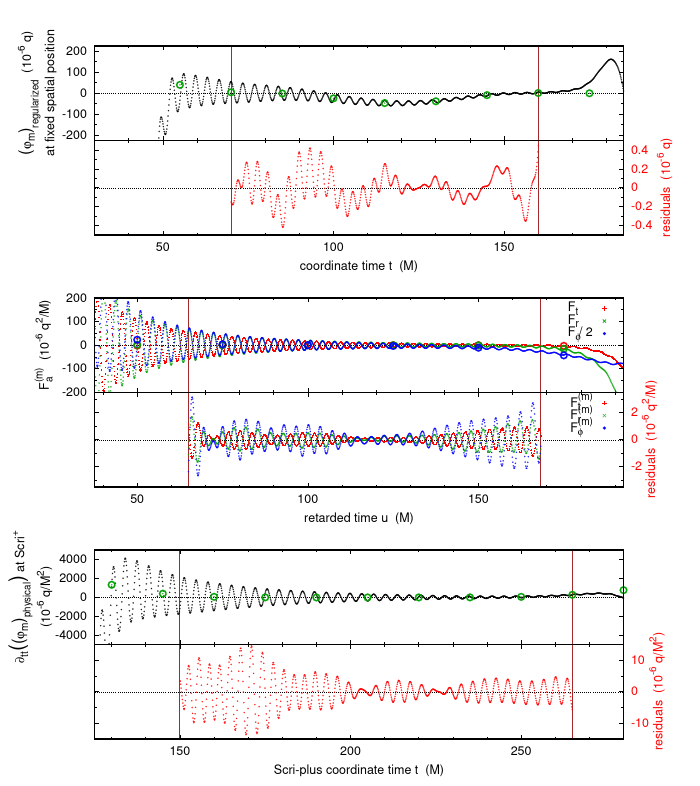}
\end{center}
\caption{
\label{fig:sf-a99p3e8-m4-fits-and-residuals}
	This figure shows the wiggles and fit residuals for the $m=4$
	scalar-field diagnostics
	for a single particle orbit ($t \in [13T_r, 14T_r]$) of the
	$(\tilde{a},p,e) = (0.99,3,0.8)$ configuration.
	In the upper subfigure, the upper plot shows the regularized
	scalar field at $r_* = 15M$
	($r = 11.810M$) as a function of coordinate time~$t$,
	at times when this position is within the worldtube,
	while the lower plot shows the residuals after fitting to
	the model~\protect\eqref{eqn:scalar-field=bg+sum-of-damped-sinusoids}.
	In the middle subfigure, the upper plot shows the 3~nontrivial
	coordinate components of the 4-vector self-force $F_a^{(m)}$
	as a function of retarded time $u$,
	while the lower plot shows the residuals after fitting to
	the model~\protect\eqref{eqn:scalar-F-a=bg+sum-of-damped-sinusoids}.
	In the lower subfigure, the upper plot shows the second time
	derivative of the physical scalar field on the equator at $\Scri^+$
	as a function of the $\Scri^+$~coordinate time~$t$,
	while the lower plot shows the residuals after fitting to the
	model~\protect\eqref{eqn:scalar-field@Scri=bg+sum-of-damped-sinusoids}.
	In each subfigure the vertical tan lines mark the interval
	over which the model is fitted, the circles in the upper subfigure
	show the background spline control points, and the differences
	between the data and the fitted model are too small to be visible
	on the scale of the upper plot.
	}
\end{figure*}

Figures~\ref{fig:sf-a99p3e8-m1-fits-and-residuals}
and~\ref{fig:sf-a99p3e8-m4-fits-and-residuals} show our model fits 
to these wiggles for $m=1$ and $m=4$ respectively.
Notice that in each case the spline control points span a wider range
of $t$ or $u$ than the range over which the model is fitted.  The
$y$~coordinates at the spline control points outside the model-fitting
range are still adjusted by the least-squares fitting algorithm, but
have only small influences on the model within the fitting range.

\begin{figure}[b]
\begin{center}
\includegraphics[width=\columnwidth]{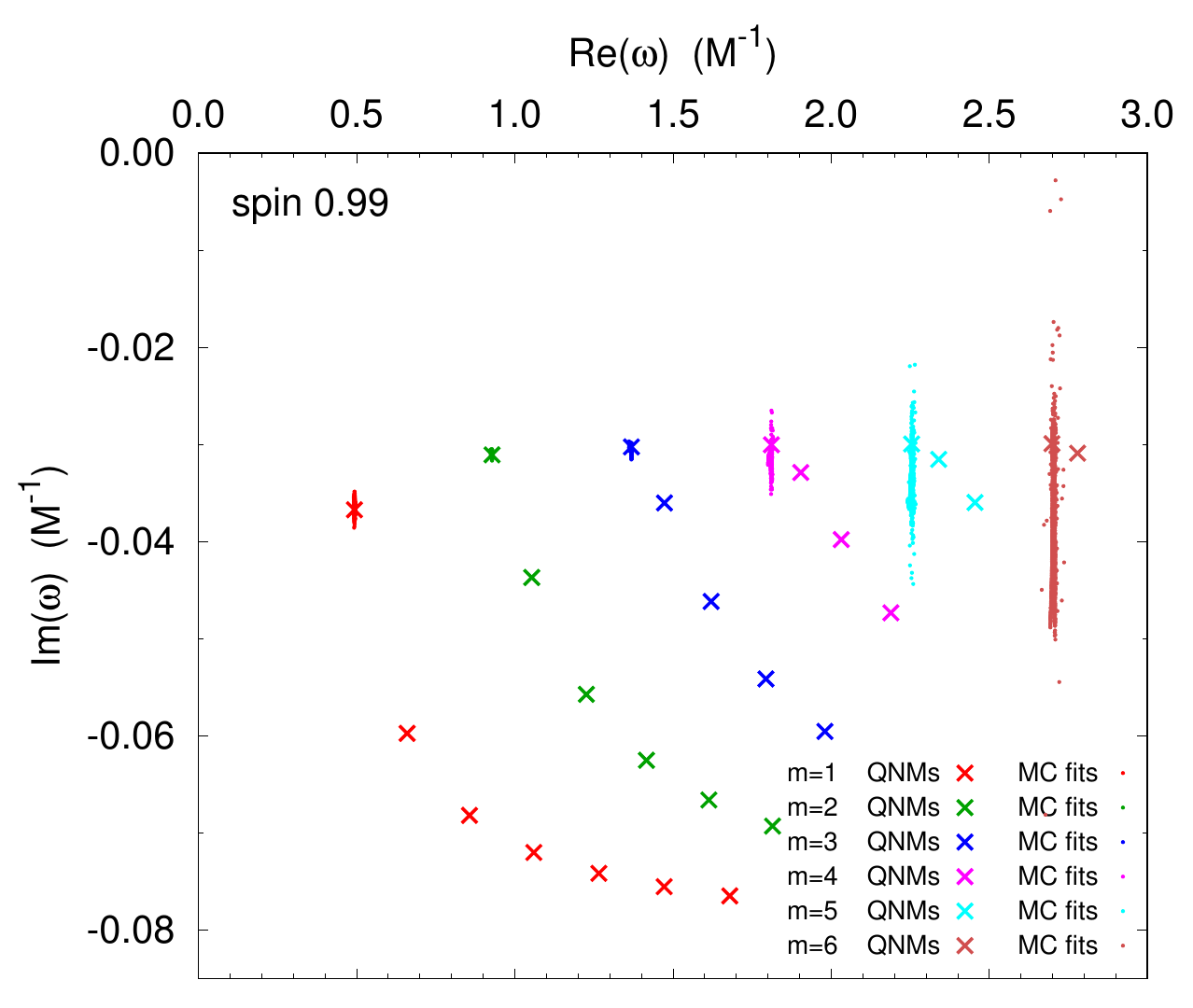}
\end{center}
\caption{
\label{fig:sf-a99p3e8-all-QNMs-and-m1-6-MC-frequencies}
	This figure shows all the Kerr QNM frequencies in the region
	$\bigl(\realpart{\omega}, \imagpart{\omega} \bigr)
		\in \bigl( [0,3] \times [-0.085,0] \bigr) M^{-1}$
	for Kerr spin~$\tilde{a} \,{=}\, 0.99$,
	together with our fitted complex frequencies'
	Monte-Carlo error estimates
	for the $m \,{=}\, 1$ through $m \,{=}\, 6$ modes of the
	$(\tilde{a},p,e) = (0.99,3,0.8)$ configuration.
	(Some of these QNM frequencies and all of the Monte-Carlo
	error estimates are also plotted at different scales in
	figure~\protect\ref{fig:sf-a99p3e8-least-damped-QNMs-and-m1-6-MC-frequencies-zoomed}.)
	The apparent anisotropy of the Monte-Carlo ``point clouds''
	in this plot is a visual illusion due to the anisotropic
	plot scale; the point clouds are actually approximately isotropic
	in $\bigl( \realpart{\omega}, \imagpart{\omega} \bigr)$.
	}
\end{figure}

\begin{figure*}[p]
\begin{center}
\includegraphics[width=\textwidth]{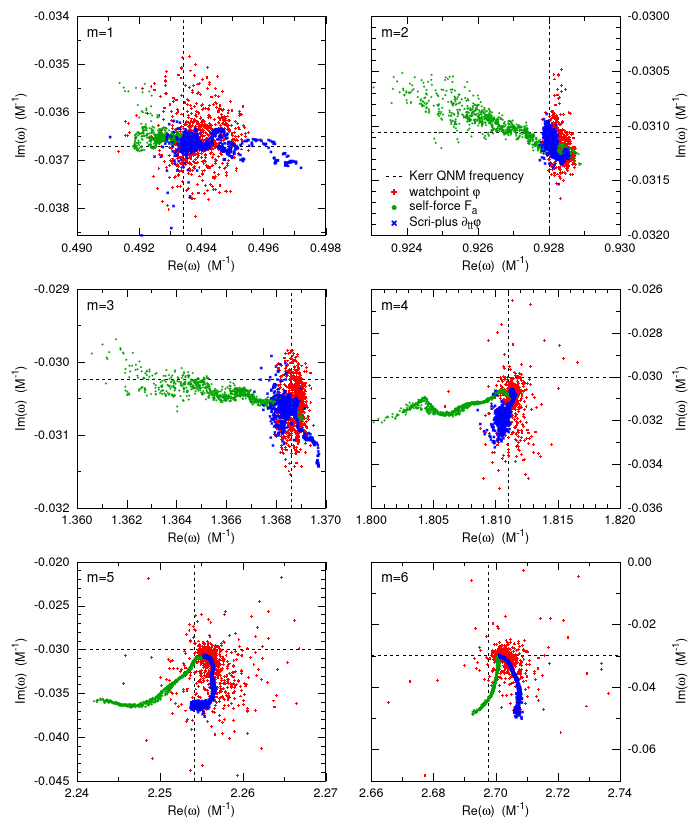}
\end{center}
\caption{
\label{fig:sf-a99p3e8-least-damped-QNMs-and-m1-6-MC-frequencies-zoomed}
	This figure shows the least-damped Kerr spin~$\tilde{a} = 0.99$
	QNM frequencies for $m=1$ through $m=6$, together with our
	fitted complex frequencies' Monte-Carlo error estimates for the
	$m=1$ through $m=6$ modes of the
	$(\tilde{a},p,e) = (0.99,3,0.8)$ configuration.
	These same Monte-Carlo error estimates
	are all plotted at a common scale in
	figure~\protect\ref{fig:sf-a99p3e8-all-QNMs-and-m1-6-MC-frequencies};
	in comparison, this figure zooms in on a small region 
	around the least-damped QNM for each~$m$.
	In each subfigure the dashed lines intersect at the least-damped
	QNM frequency.  The legend is common to all of the subfigures.
	}
\end{figure*}

\begin{figure*}[p]
\begin{center}
\includegraphics[width=\textwidth]{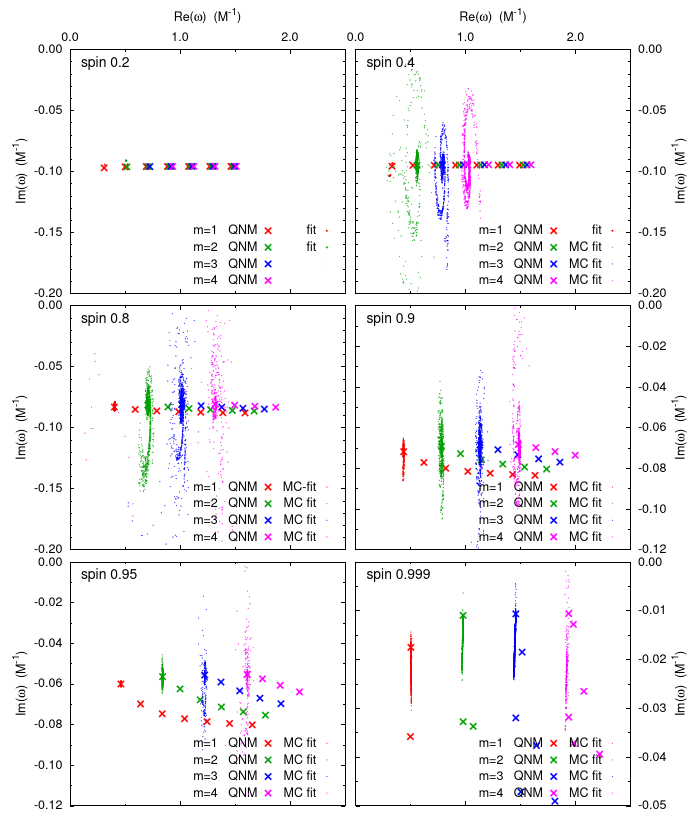}
\end{center}
\caption{
\label{fig:sf-other-spins-all-QNMs-and-m1-4-MC-frequencies}
	This figure shows all the Kerr QNM frequencies in the
	plot regions,
	together with our fitted complex frequencies' Monte-Carlo
	error estimates
	for the $m \,{=}\, 1$ through $m \,{=}\, 4$ modes
	of the all the Kerr spin~$\tilde{a} \,{\ne}\, 0.99$ configurations.
	(This is analogous to
	figure~\protect\ref{fig:sf-a99p3e8-all-QNMs-and-m1-6-MC-frequencies}
	for $\tilde{a} \,{\ne}\, 0.99$.)
	The apparent anisotropy of the Monte-Carlo ``point clouds''
	in this plot is a visual illusion due to the anisotropic
	plot scales; the point clouds are actually approximately isotropic
	in $\bigl( \realpart{\omega}, \imagpart{\omega} \bigr)$.
	}
\end{figure*}

Figures~\ref{fig:sf-a99p3e8-all-QNMs-and-m1-6-MC-frequencies}
--\ref{fig:sf-other-spins-all-QNMs-and-m1-4-MC-frequencies}
show the fitted complex frequencies and their Monte-Carlo error estimates,
compared to Kerr QNM frequencies calculated by Berti, Cardoso, and
Starinets~\cite{Berti-Cardoso-Starinets-2009:BH-etal-QNM-review,Berti06b}.
\footnote{
	 Data tables downloaded from
	 \hbox{\url{https://pages.jh.edu/~eberti2/ringdown/}}
	 on 19 April 2019.
	 }
{}
In each case the fitted frequencies agree with the calculated QNM
frequencies, lending further support to the identification of wiggles
with QNMs (more precisely, QNMs sampled at the observation points).

\begin{figure*}[tp]
\includegraphics[width=\textwidth]{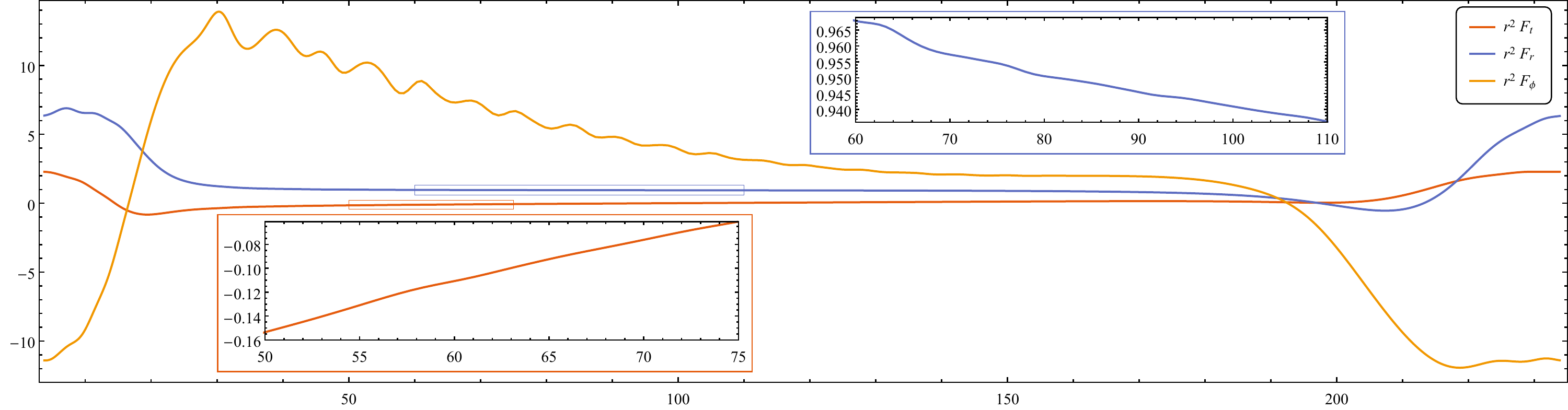}
\includegraphics[width=\textwidth]{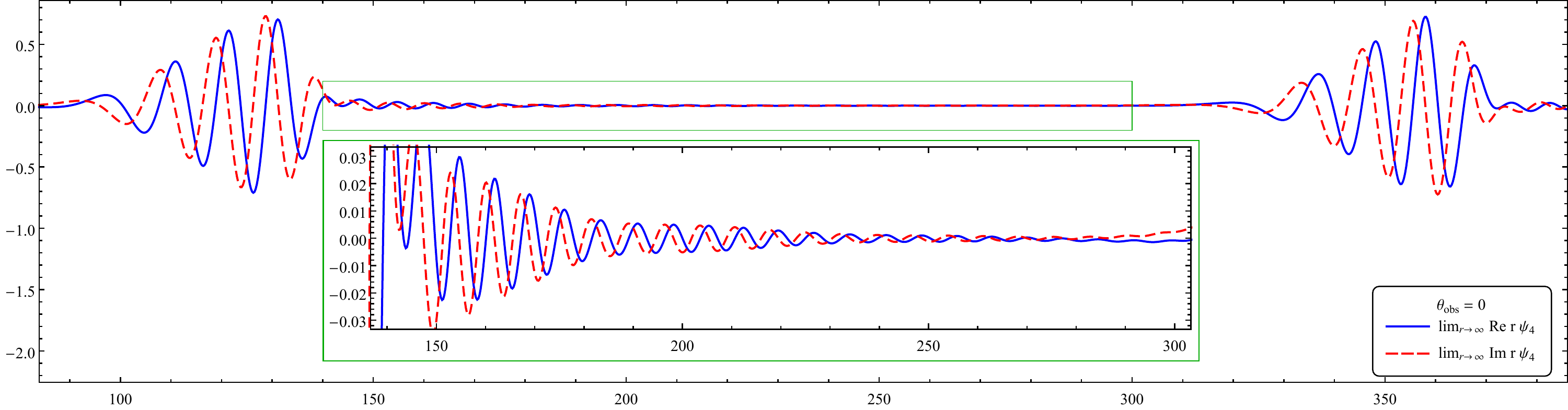}
\includegraphics[width=\textwidth]{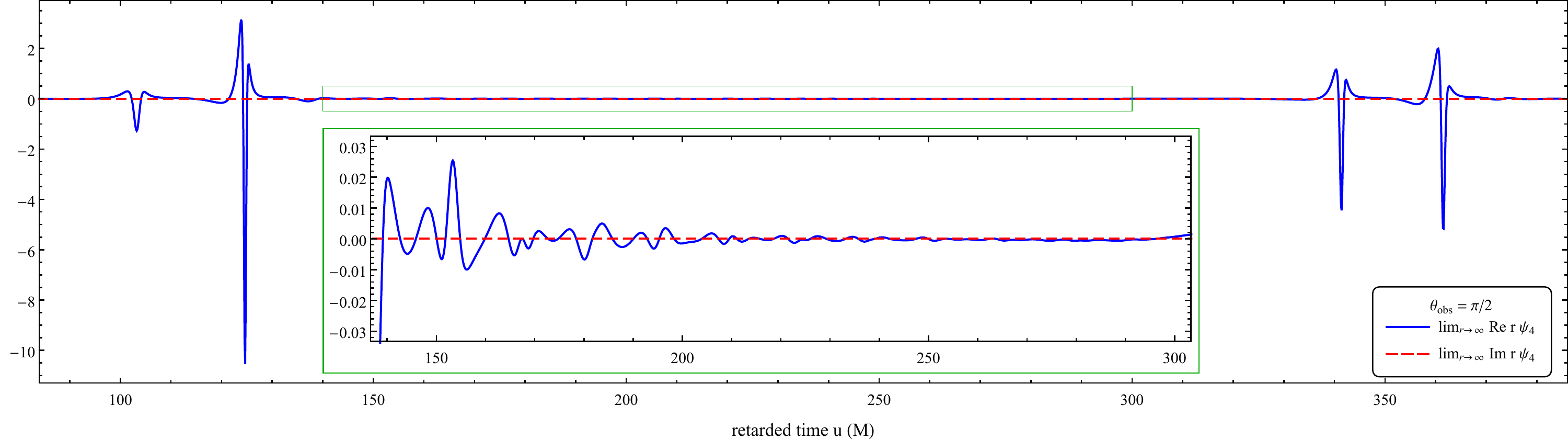}
\caption{
\label{fig:grava99p3e8-full}
	Three different aspects of the gravitational perturbation
	for the $(\tilde{a},p,e) = (0.99,3,0.8)$ configuration.
	\emph{(top)}
		The three non-zero components of the gravitational
		self-force at the particle. The $F_{\phi}$~component
		has very prominent wiggles. The wiggles in the other
		two components are present but only visible after zooming in. 
	\emph{(middle)}
		The value of $\psi_4$ observed at $\Scri^+$ when the
		system is viewed ``face on'' ($\theta_\observer=0$).
		We observe very clean exponentially decaying wiggles
		right up to the next burst produced when the particle
		approaches the central black hole again.
	\emph{(bottom)}
		The value of $\psi_4$ observed at $\Scri^+$ when the
		system is viewed ``edge on'' ($\theta_\observer=\pi/2$).
		In this case the wiggle pattern is much more complex
		as multiple $(\ssl,m)$-modes contribute.}
\end{figure*}

\begin{figure*}[tp]
\includegraphics[width=\textwidth]{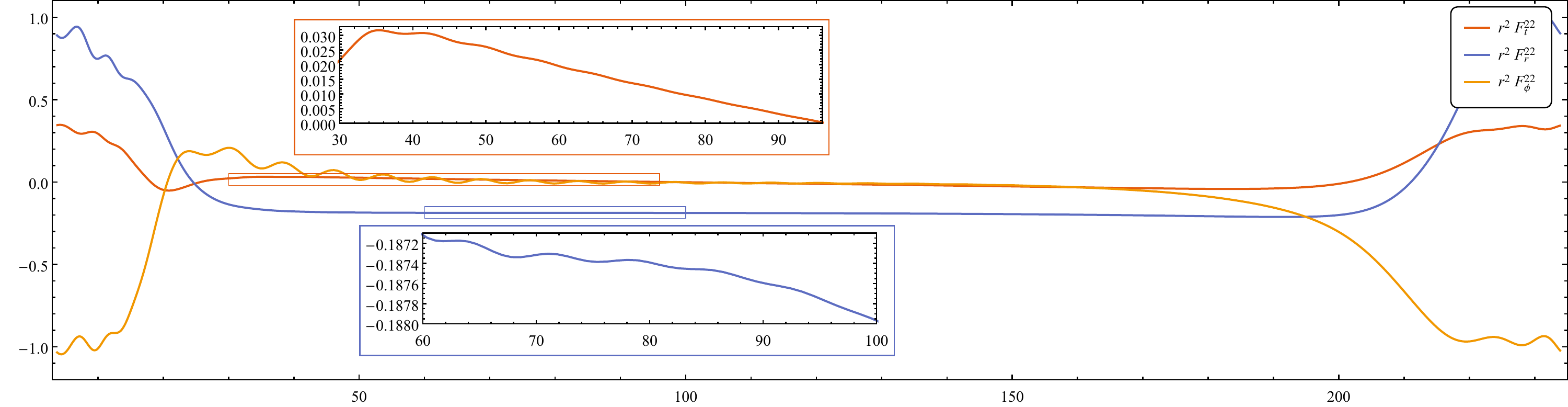}
\includegraphics[width=\textwidth]{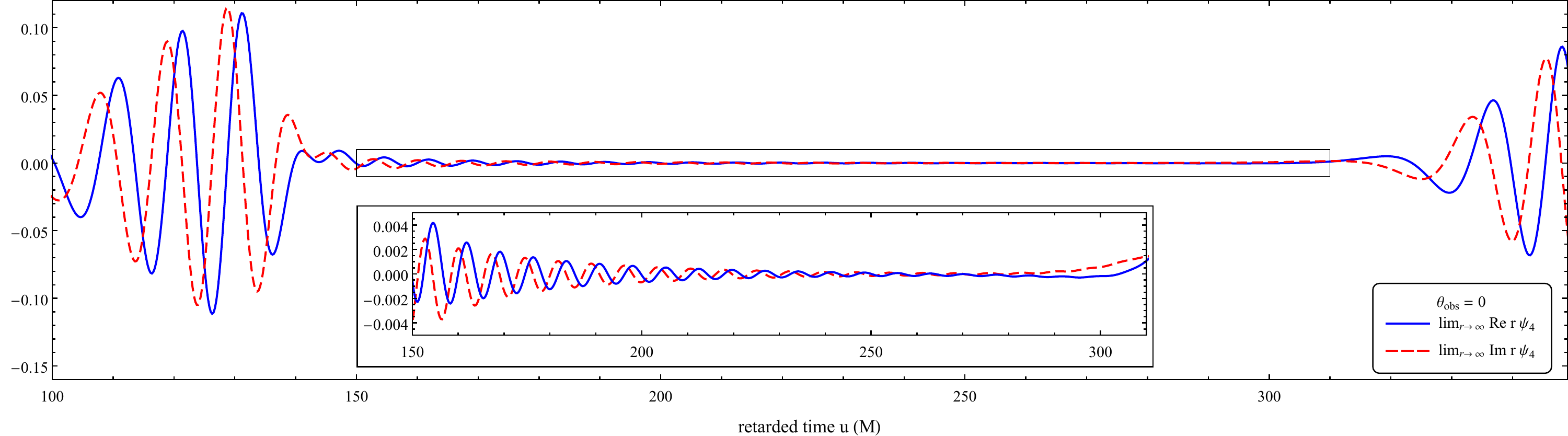}
\includegraphics[width=\textwidth]{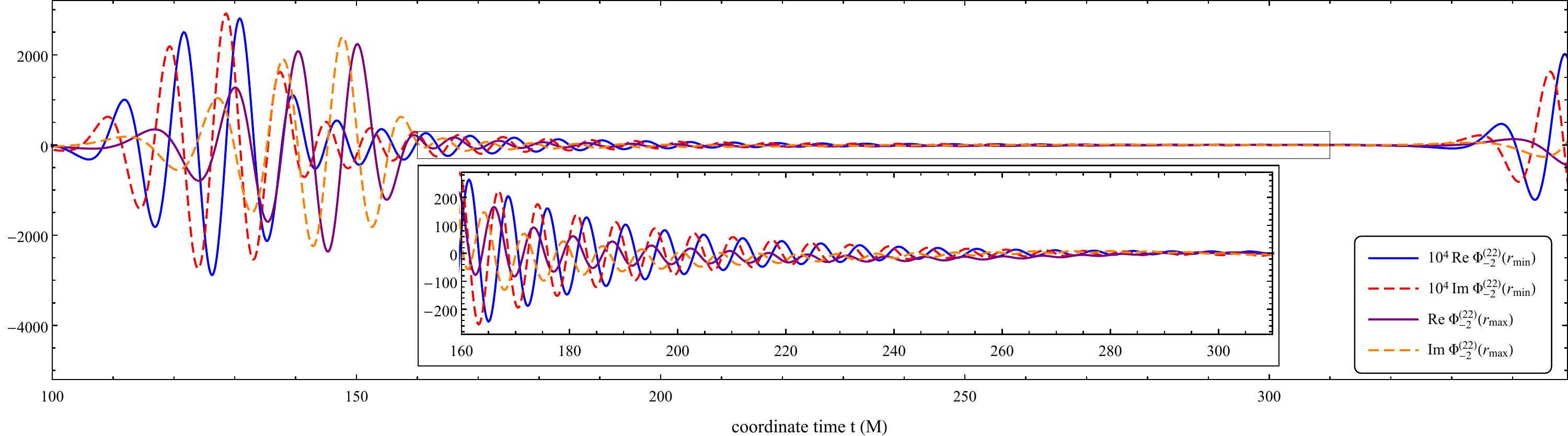}
\caption{
\label{fig:grava99p3e8-l2m2}
	In this figure we focus on the $\ssl=m=2$ mode of the
	gravitational perturbation 
	for the $(\tilde{a},p,e) = (0.99,3,0.8)$ configuration.
	\emph{(top)}
		The three non-zero components of the gravitational self-force.
	\emph{(middle)}
		The value of $\psi_4$ observed at $\Scri^+$ when
		the system is viewed ``face on'' ($\theta_\observer=0$).
	\emph{(bottom)}
		The Teukolsky variable $\Phi_{-2}^{(22)}$ evaluated
		on the symmetry axis $\theta = 0$ and at
		two different radii $r=r_{\max}$ and $r=r_{\min}$.  	}
\end{figure*}



\subsection{Gravitational field}
\label{sect:data-and-QNM-fits/gravitational-field}

We now turn our attention to gravitational perturbations.
We first consider the same $(\tilde{a},p,e) = (0.99, 3, 0.8)$
configuration studied in figures~\ref{fig:sf-a99p3e8-m1-wiggles-overview}--\ref{fig:sf-a99p3e8-least-damped-QNMs-and-m1-6-MC-frequencies-zoomed} in the scalar field case. Figure~\ref{fig:grava99p3e8-full} displays both the gravitational self-force at the particle location and the waveform observed at $\Scri^+$. When looking at the local self-force the wiggles are most pronounced in the $F_\phi$ component. However, faint traces of wiggles can be found by zooming in on the $F_t$ and $F_r$ components. We note that the relative amplitudes of the wiggles in the gravitational self-force are much smaller than those in the scalar case for the same orbit
(shown in figures~\ref{fig:sf-a99p3e8-m1-wiggles-overview}
and~\ref{fig:sf-a99p3e8-m4-wiggles-overview}).

The waveform observed at $\Scri^+$ depends on the viewing angle. When the system is viewed ``face on'' (middle panel of figure~\ref{fig:grava99p3e8-full}) the waveform is determined by the $m=2$ modes with the $m=\ssl=2$ dominating. In this case the wiggles appear as a clear exponentially damped sinusoid. When the system is viewed ``edge on'' (bottom panel of figure~\ref{fig:grava99p3e8-full}), the wiggles have a much more irregular shape, consistent with a much larger collection of $m$'s and $\ssl$'s contributing to the wiggles. Also note that while the overall waveform has a much larger amplitude when viewed edge on (due to contributions from higher modes), the observed wiggles are actually stronger when the system is viewed ``face on''. This is consistent with the wiggles being dominated by the $\ssl=m=2$ mode.

One of the advantages of using a frequency domain approach is that we can easily isolate individual $\ssl m$-modes (as defined in section~\ref{sect:calculations-of-Kerr-perturbations/gravitational}). Figure~\ref{fig:grava99p3e8-l2m2} shows different aspects of the $\ssl=m=2$ mode of the gravitational perturbation generated by a particle on our standard $(\tilde{a},p,e) = (0.99, 3, 0.8)$ configuration. The $\ssl=m=2$ mode of the gravitational self-force (top panel) shows the same qualitative features as the full GSF; the $F_\phi$ components show the most obvious wiggles with weak wiggles visible in the other components. The $\ssl=m=2$ mode of the field observed at  $\Scri^+$ shows a clean exponentially decaying sinusoid wiggle just as the full field. In addition, the bottom panel of figure~\ref{fig:grava99p3e8-l2m2} shows the local Teukolsky variable $\Phi^{(22)}_{-2}$
at two watch points located on the background Kerr spacetime's symmetry axis
at radii corresponding to the periapsis and apoapsis of the particle orbit.
These show the cleanest wiggles of any of our diagnostics.

To test our hypothesis that the observed wiggles are, in fact, QNM
excitations, we perform a global fit of our three field diagnostics
(local gravitational self-force, field at $\Scri^+$, and field at
watchpoints) following the methodology set out in
section~\ref{sect:QNM-models-and-fits/gravitational}.
Table~\ref{tab:gravqnmfits} summarizes the results for some low
order $\ssl m$-modes.  In each, case we recover the principal QNM
frequency and damping time of the gravitational field within the
estimated numerical precision of the fits.  This provides yet more
evidence for our hypothesis that the observed wiggles are QNM
excitations.  Note that while our fits include multiple QNMs, we
do not conclusively recover any of the modes beyond the principal
mode.  (More precisely, we find that the estimated numerical errors of the
recovered complex frequencies are comparable to the variation of the initial
seed for the optimization.)  Not including the higher modes, however, led to
observable bias in the recovery of the principal QNMs.

\begin{table}[t]
\caption{
\label{tab:gravqnmfits}
	Numerical fits of the wiggles in the gravitational
	(spheroidal) $(\ssl m)$-mode field diagnostics
	for the $(\tilde{a},p,e) = (0.99,3,0.8)$ configuration.
	In each row the top values are the fitted frequency/decay rate;
	the bottom values are the known values for (least damped)
	gravitational QNMs.  Fitting errors are indicated in parentheses,
	\eg{}, $0.870891(7)$ means $0.870891\pm 7\times10^{-6}$.}
\begin{ruledtabular}
\begin{tabular}{ccll}
%
		\text{$\ssl$} & \text{$m$} & \text{$\omega_{\ssl mk} $} & \text{$\alpha_{\ssl mk}$ }\\
		\hline
 2 & 0 & 
\begin{tabular}{l}
 0.421(3) \\
 0.423 \\
\end{tabular}
 & 
\begin{tabular}{l}
 0.069(4) \\
 0.072 \\
\end{tabular}
 \tst\\
 \hline
 2 & 1 & 
\begin{tabular}{l}
 0.57278(6) \\
 0.57274 \\
\end{tabular}
 & 
\begin{tabular}{l}
 0.04624(5) \\
 0.04620 \\
\end{tabular}
 \tst\\
  3 & 1 & 
\begin{tabular}{l}
 0.7891(1) \\
 0.7891 \\
\end{tabular}
 & 
\begin{tabular}{l}
 0.0618(3) \\
 0.0618 \\
\end{tabular}
 \tst\\
 \hline
 2 & 2 & 
\begin{tabular}{l}
 0.870891(7) \\
 0.870892 \\
\end{tabular}
 & 
\begin{tabular}{l}
 0.029386(4) \\
 0.029390 \\
\end{tabular}
 \tst\\
 3 & 2 & 
\begin{tabular}{l}
 0.999142(9) \\
 0.999140 \\
\end{tabular}
 & 
\begin{tabular}{l}
 0.037128(6) \\
 0.037129 \\
\end{tabular}
 \tst\\
 4 & 2 & 
\begin{tabular}{l}
 1.1771(1) \\
 1.1770 \\
\end{tabular}
 & 
\begin{tabular}{l}
 0.0510(4) \\
 0.0511 \\
\end{tabular}
 \tst\\
 \hline
 3 & 3 & 
\begin{tabular}{l}
 1.32308(3) \\
 1.32308 \\
\end{tabular}
 & 
\begin{tabular}{l}
 0.02940(1) \\
 0.02940 \\
\end{tabular}
 \tst\\
 4 & 3 & 
\begin{tabular}{l}
 1.43371(3) \\
 1.43372 \\
\end{tabular}
 & 
\begin{tabular}{l}
 0.03323(3) \\
 0.03323 \\
\end{tabular}
 \tst\\
 5 & 3 & 
\begin{tabular}{l}
 1.58353(3) \\
 1.58351 \\
\end{tabular}
 & 
\begin{tabular}{l}
 0.04306(3) \\
 0.04308 \\
\end{tabular}
 \tst\\
 \hline
 4 & 4 & 
\begin{tabular}{l}
 1.7731(2) \\
 1.7731 \\
\end{tabular}
 & 
\begin{tabular}{l}
 0.02946(6) \\
 0.02947 \\
\end{tabular}
 \tst\\
 5 & 4 & 
\begin{tabular}{l}
 1.87254(4) \\
 1.87254 \\
\end{tabular}
 & 
\begin{tabular}{l}
 0.03160(5) \\
 0.03160 \\
\end{tabular}
 \tst\\
 \hline
 5 & 5 & 
\begin{tabular}{l}
 2.2216(2) \\
 2.2217 \\
\end{tabular}
 & 
\begin{tabular}{l}
 0.0295(1) \\
 0.0295 \\
\end{tabular}
 \tst\\

\end{tabular}
\end{ruledtabular}
\end{table}

\subsection{Dependence on orbital parameters}
\begin{figure}[tb!]
\includegraphics[width=\columnwidth]{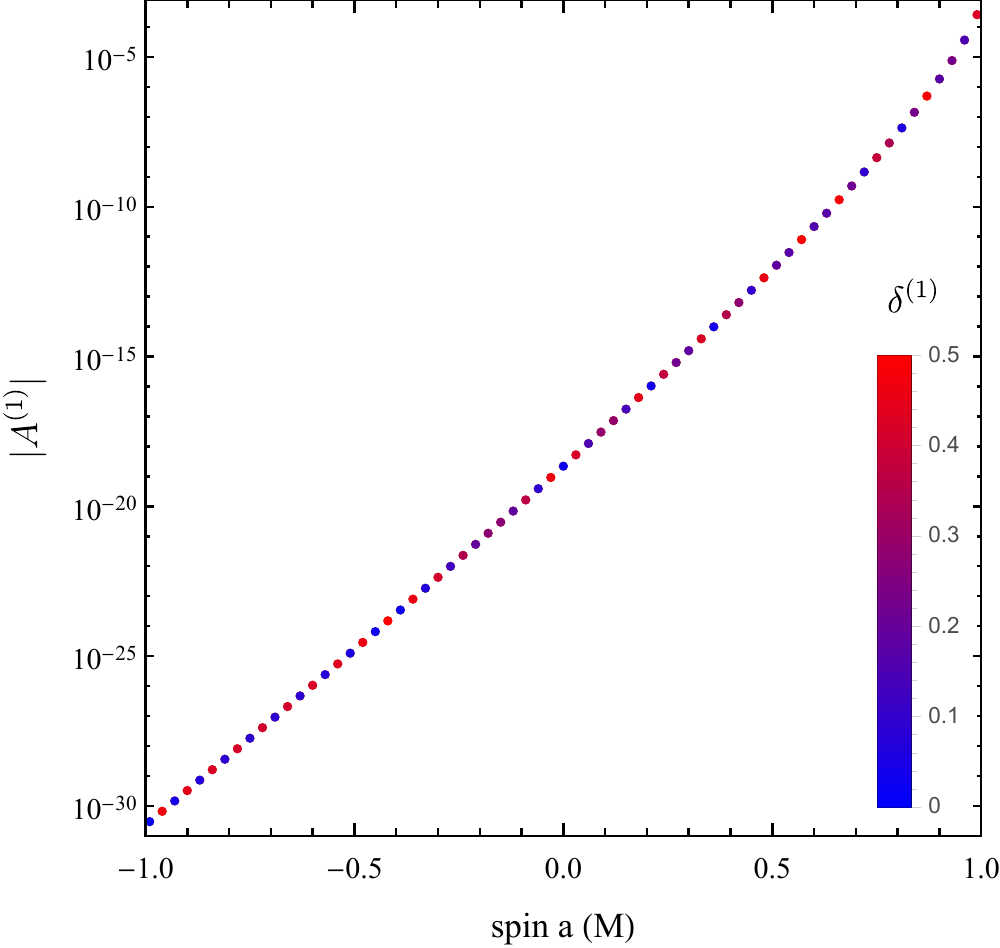}
\caption{
\label{fig:spinseries}
	Dependence of the the fitted QNM amplitude~$|A_1|$
	of the lowest damped $\ssl=m=2$ QNM as a function of the
	primary spin~$a$ for a sequence of orbits with fixed $e=0.8$ and ratio
	$\Omega_\phi/\Omega_r = 2.684\,694\,379\dots$
	(the frequency ratio for the
	$(\tilde{a},p,e) = (0.99,3,0.8)$ configuration).
	The parameter $u_\rref$ is fixed to coincide with the
	particle passing through apoapsis.  The data points are
	shaded according to the degree of alignment $\delta$ of
	the particle spectrum with the QNM frequency~$\omega^{(1)}$.}
\end{figure}

\begin{figure}[tb]
\includegraphics[width=\columnwidth]{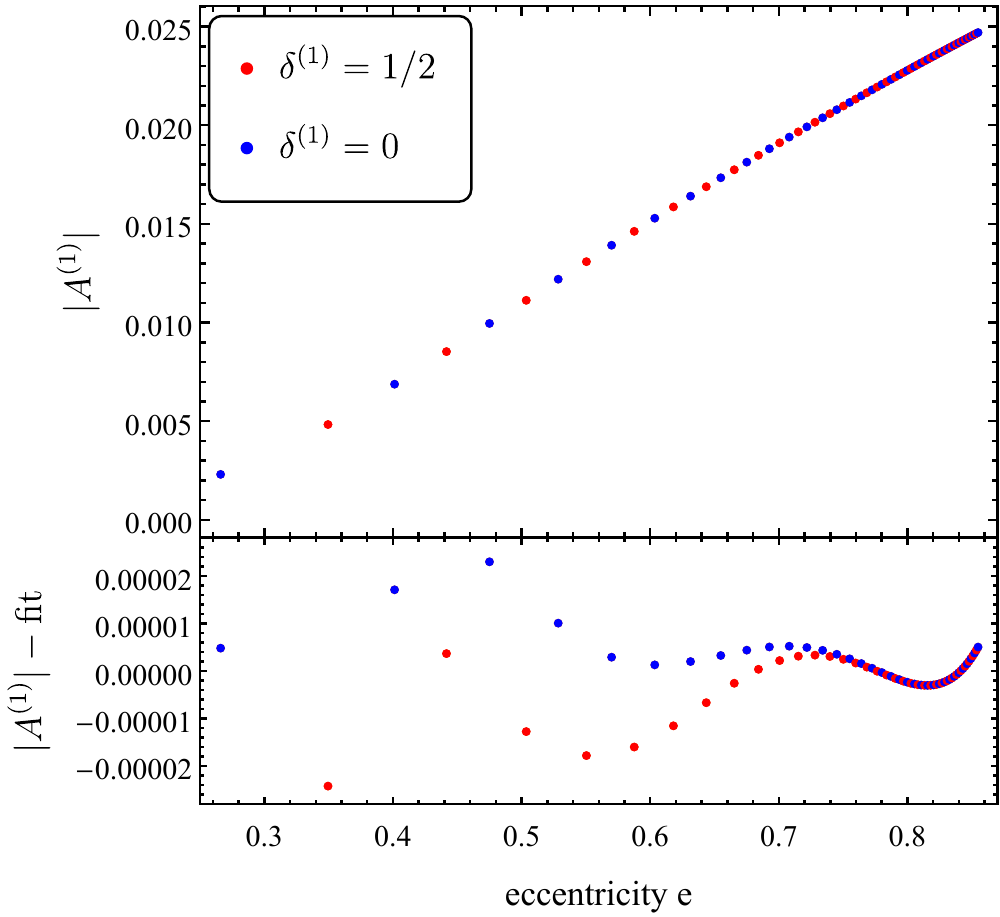}
\caption{
\label{fig:eccseries}
	Dependence of the the fitted QNM amplitude~$|A_1|$
	of the lowest damped $\ssl=m=2$ QNM as a function of the
	eccentricity~$e$ for a sequence of orbits with fixed
	primary spin~$a=0.95M$ and periapsis distance
	$r_{\min} = 1.85M$.  The parameter $u_\rref$
	is fixed to coincide with the particle passing
	through periapsis.  The data points are shaded
	according to the degree of alignment $\delta$ of the
	particle spectrum with the QNM frequency~$\omega^{(1)}$.
	The lower panel shows the difference between the
	amplitude and a fitted quintic polynomial in~$e$.}
\end{figure}

\begin{figure}[tbh]
\includegraphics[width=\columnwidth]{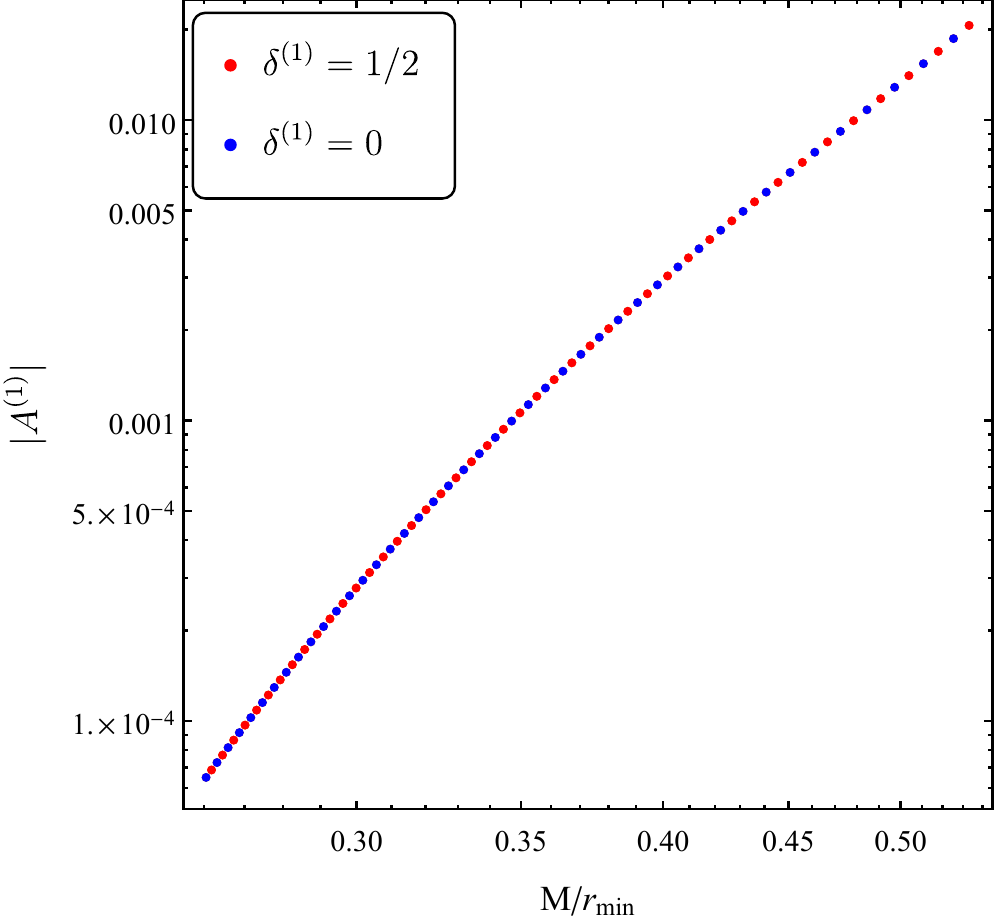}
\caption{
\label{fig:paseries}
	Dependence of the the fitted QNM amplitude~$|A_1|$
	of the lowest damped $\ssl=m=2$ QNM as a function
	of the inverse periapsis distance~$M/r_{\min}$ for a
	sequence of orbits with fixed primary spin~$a=0.95M$
	and eccentricity~$e = 0.8$. The parameter $u_\rref$
	is fixed to coincide with the particle passing through
	periapsis.  The data points are shaded according to the
	degree of alignment $\delta$ of the particle spectrum
	with the QNM frequency~$\omega^{(1)}$.}
\end{figure}

In this section we study how the strength of the QNM excitations in
the gravitational field depends on the parameters of the orbit.  For this
investigation we leverage the ease with which the $(\ssl m)$~modes of the
gravitational field can be computed using our frequency domain code.  As
a measure of the strength of the excitations we take the amplitudes~$A^{(k)}_c$
and~$A^{(k)}_s$ in~\eqref{eq:gravscriplusmodel}, which we combine to define
\begin{eqnarray}
\abs{A^{(k)}} :=\sqrt{\hh{A^{(k)}_c}^{2}+\hh{A^{(k)}_s}^2}.
\end{eqnarray}
For the purpose of this investigation we assume that the observed wiggles
are indeed QNM excitations, we therefore determine the $\abs{A^{(k)}}$
by a linear fit, keeping the frequencies and decay rates fixed at the
exact QNM values.  The value of $\abs{A^{(k)}}$ depends on the choice
of $u_\rref$ in~\eqref{eq:gravscriplusmodel}.  In this section we choose
$u_\rref$ to coincide with the particle passing through either the
periapsis or apoapsis of the orbit.

We are particularly interested in whether excitation of the QNMs exhibits
a strong dependence on the alignment of the discrete orbital frequency
spectrum of the orbit with the QNM mode.  If strong localized ``resonances''
between the orbit frequency spectrum and QNM excitations were to exist,
these could have significant impact on waveform modelling strategies,
as they would hamper an attempt to apply reduced order modelling to
build efficient waveforms.  On the other hand, such a phenomenon might
lead to interesting and rich dynamics.
To quantify the alignment between a QNM and the particle orbit we
define
\begin{equation}
\delta^{(k)} :=
	\min_{n\in\ZZ}
	\frac{\abs{\omega^{(k)} - m \Omega_\phi - n\Omega_r}}{\Omega_r},
\end{equation}
i.e., $\delta^{(k)}$ is the distance between the QNM frequency~$\omega^{(k)}$
and the nearest line in the orbit's frequency spectrum, normalized such
that $\delta^{(k)} \,{=}\, 0$ corresponds to maximal alignment
and $\delta^{(k)} \,{=}\, 1/2$ to maximal misalignment.

In figure~\ref{fig:spinseries} we examine the dependence of the QNM amplitude
(for the least-damped ($k=1$) $\ssl=m=2$ QNM) on the spin~$a$ of the
background Kerr spacetime while keeping eccentricity and ratio of the orbital frequencies (matching the eccentricity and frequency ratio for the $(\tilde{a},p,e) = (0.99,3,0.8)$ configuration).
We see that the dependence of the QNM excitation amplitude
on the Kerr spin~$a$ is very smooth, with no noticeable dependence on
the spectrum misalignment parameter~$\delta$. Notice that the QNM
excitations persist for negative spins (i.e., retrograde orbits),
although they become exceedingly weak.

Figure~\ref{fig:eccseries} explores the dependence of the amplitude~$\abs{A^{(1)}}$ of the least damped ($k=1$) $\ssl=m=2$ QNM on the particle's
orbital eccentricity.  For this exploration we keep the spin of the
background Kerr spacetime fixed at $a=0.95M$, and we fix the periapsis
distance at $r_{\min} = 1.85M$.  The relationship between $\abs{A^{(1)}}$
and $e$ appears almost linear by eye, with slight deviations both at high
and low eccentricity.  If we subtract off the dominant trend in the form
of a quintic fit in $e$, we see what appears to be a systematic trend
where orbits with $\delta^{(1)}=0$ have a slightly larger
amplitude~$\abs{A^{(1)}}$ than orbits with $\delta^{(1)}=1/2$.
This difference becomes stronger for low eccentricity orbits.
This latter effect
is consistent with the frequency spectrum of the orbit becoming
sparser at lower eccentricities.  However, we stress that this effect
is very small, with the variation of the QNM amplitude~$\abs{A^{(1)}}$
due to changing~$\delta^{(1)}$ being only about 1~part in $10^3$. 

Finally, figure~\ref{fig:paseries} explores the relation between the
amplitude~$\abs{A^{(1)}}$ of the least damped ($k=1$) $\ssl=m=2$ QNM
and the particle's inverse periapsis distance $M/ r_{\min}$, keeping
the Kerr spin ($a = 0.95 M$) and particle eccentricity ($e = 0.8$) fixed.
As is to be expected, the amplitude~$\abs{A^{(1)}}$ drops off sharply
as we increase the particle periapsis radius.

We emphasize that the overall shape of the plots in
figures~\ref{fig:spinseries}--\ref{fig:paseries}
depends sensitively on the choice of $u_\rref$, hence one should not
read too much into the shapes themselves.  However, there are three
main lessons that we learn from this investigation that do not depend
on the choice of~$u_\rref$:
\begin{itemize}
\item   The amplitudes of the wiggles depend smoothly on the Kerr
	spin (figure~\ref{fig:spinseries}) and orbital parameters
	(figures~\ref{fig:eccseries} and~\ref{fig:paseries}).
	In particular, no fine-tuning is needed for wiggles to appear.
\item	The wiggles are strongest for high spin
	and prograde particle orbits with high eccentricity
	and low periapsis distance.  However, there is no indication
	that they will completely disappear in any region of the
	parameter space (although they may become very difficult
	to separate from the rest of the field due to low amplitudes,
	high damping rates, and/or longer periods).
\item	The effect of aligning the orbital frequencies with the QNM
	frequencies is very small, and decreases still further when
	the orbital spectrum becomes denser for more eccentric orbits.
\end{itemize}


\section{Discussion and Conclusions}
\label{sect:discussion-and-conclusions}

\begin{figure}[tbp]
\includegraphics[width=\columnwidth]{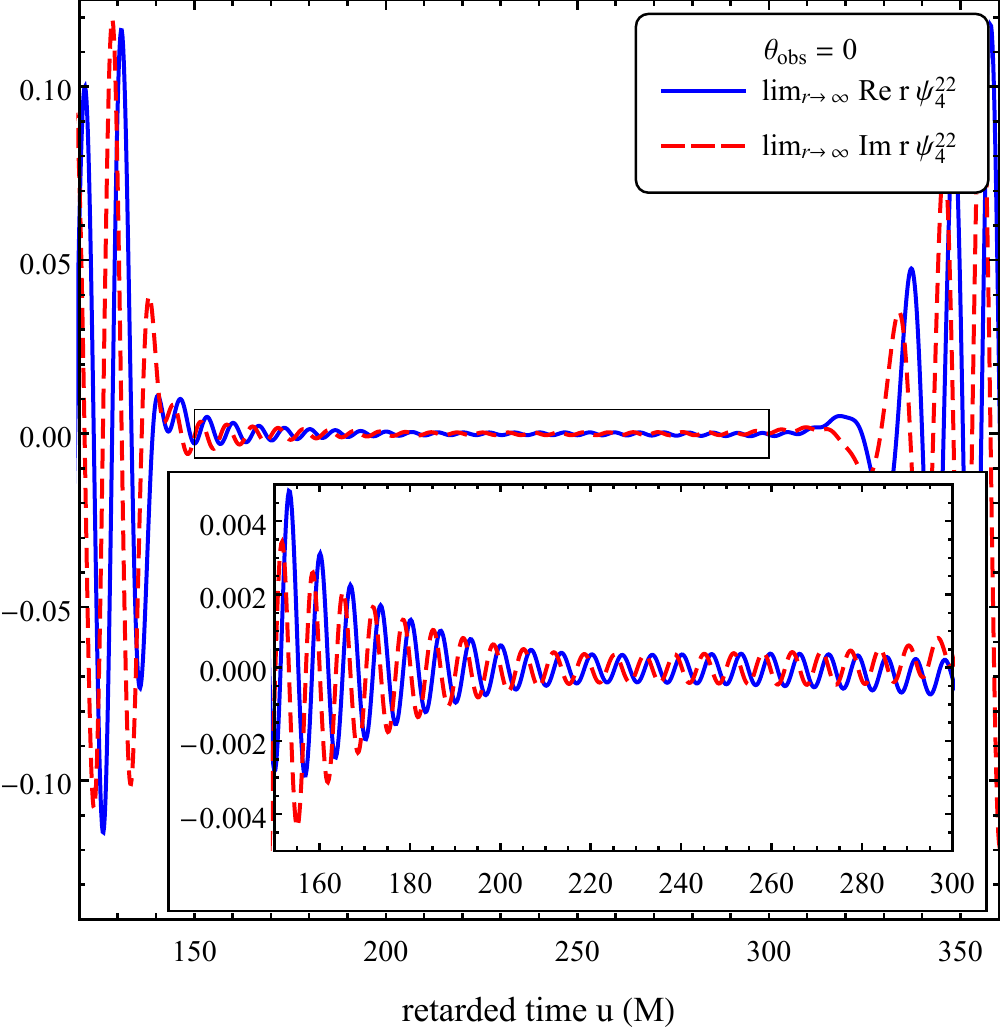}
\caption{
\label{fig:WTF}
	The $\ssl=m=2$ mode of the gravitational waveform
	at $\Scri^{+}$ for the configuration
	$(\tilde{a},p,e) = (0.999\,99,2.918,0.807)$.
	The observed wiggles are surprising because the real part
	of the frequency (in the highlighted area) lies between
	$0.93 M^{-1}$ and $0.98 M^{-1}$, whereas the QNM frequencies
	are bunched up near $0.995 M^{-1}$.}
\end{figure}

In this paper we study an interesting class of features first observed
in the scalar self-force for point particles in orbit in Kerr
spacetime~\cite{Thornburg-Wardell-2017:Kerr-scalar-self-force}.
That study identified the feature, introduced the term ``wiggles'',
and argued that it was in some (unspecified) manner
``\emph{caused} by the particle's close passage by the large black hole'',
but did not attempt to attribute it to any particular physical origin.
More recently Refs.~\cite{Thornburg-2016-Capra-Meudon-talk,
Thornburg-2017-Capra-Chapel-Hill-talk,
Nasipak-Osburn-Evans-2019:Kerr-scalar-self-force-and-wiggles}
have shown further examples of wiggles, demonstrated that wiggles'
complex frequencies agree with known Kerr quasinormal-mode (QNM)
frequencies, and concluded that wiggles are in fact ``just'' a sampling
at the measurement point(s) of Kerr QNMs excited by the particle.

Here we survey the phenomenology of wiggles for both the scalar-field
and gravitational cases, across a range of Kerr spins and particle orbits.
In both the scalar-field and gravitational cases we find that wiggles
are essentially a generic phenomenon, i.e., they occur over a wide range
of configuration space without any ``fine-tuning'' of parameters.  Wiggles
are observable in field perturbations at fixed spatial positions,
in the radiation-reaction ``self-force'', and in the radiated fields
at $\Scri^+$.

In both the scalar-field and gravitational cases we find that at all
observed locations in spacetime, wiggles can be quantitatively fit by
models of QNMs sampled at the observation points.  In particular,
in both the scalar-field and gravitational cases our fitted
wiggle frequencies agree well [in both real (oscillatory) and
imaginary (damping) parts] with Kerr QNM frequencies calculated
by Berti, Cardoso, and
Starinets~\cite{Berti-Cardoso-Starinets-2009:BH-etal-QNM-review,Berti06b}
(figures~\ref{fig:sf-a99p3e8-all-QNMs-and-m1-6-MC-frequencies}
--\ref{fig:sf-other-spins-all-QNMs-and-m1-4-MC-frequencies}
and table~\ref{tab:gravqnmfits}).

The appearance of pronounced wiggles appears to rely on three key
aspects of the configuration of the system:
\begin{itemize}
\item	A highly spinning central (Kerr) black hole
	(the closer to $\tilde{a}=1$, the more pronounced the effect).
\item	A highly eccentric prograde orbit for the particle
	(the closer to $e=1$, the more pronounced the effect).
\item	A close periapsis passage by the particle
	(the closer to the light ring, the more pronounced the effect).
\footnote{
	 We refer here to the \emph{prograde} light ring; we observe only
	 very small QNM excitation when the particle
	 periapsis is close to the retrograde
	 light ring.
	 }
\end{itemize}
This is not surprising: highly spinning black holes have much
longer-lived QNMs than those with low spin. Increasing eccentricity
of the particle orbit does three things: it increases the strength of the
perturbation at the periapsis, it widens the frequency spectrum of the
perturbation (increasing the overlap with the QNMs), and it provides a natural
``quiet'' period when the particle approaches its apoapsis, during which
the QNMs can more easily be observed.  Finally, bringing the particle
periapsis closer to the light ring allows the perturbation to deposit
more energy in the QNMs.  (QNMs in Kerr spacetime are readily excited by orbits
near the light ring~\cite{Goebel-1972:BH-QNM-as-GWs-at-light-ring,
Berti-Cardoso-Starinets-2009:BH-etal-QNM-review,Khanna-Price-2017}.)

Interestingly, we find that the amplitude with which wiggles
are excited does \emph{not} depend sensitively on the particle's precise
orbital motion near periapsis.  Notably, we find that the wiggle amplitude
varies smoothly and monotonically with the particle periapsis radius
and orbital eccentricity (and with the Kerr spin).

We have not attempted to carefully delineate the exact boundaries
of the region in configuration space where wiggles occur (even assuming
that there are, in fact, configurations with \emph{no} QNM excitation,
which is not obvious).  It is likely that different modelling/fitting
schemes could observe and fit low-amplitude and/or rapidly-damped
wiggles even in some cases where we fail to observe them
(e.g., the \WigglesNo{} cases in table~\ref{tab:sf-configurations}).
For example, figure~\ref{fig:spinseries} strongly suggests that although
the wiggle amplitude is very small in some cases, wiggles are present for
\emph{all} Kerr spins along this sequence of orbits, including retrograde
as well as prograde orbits.

Our scalar-field wiggle modelling/fitting scheme is (deliberately)
quite conservative in requiring visual observation of a wiggle in a
time-series plot of the original diagnostic.  This requirement reduces
the risk of false positives (where we would misidentify a fitting or
background-spline artifact as a wiggle), at the cost of reducing our
sensitivity to low-amplitude and/or rapidly-damped wiggles.

An interesting example of these factors at play is the
$(\tilde{a},p,e) = (0.99,8,0.8)$ scalar-field configuration,
for which Nasipak, Osburn, and
Evans~\cite{Nasipak-Osburn-Evans-2019:Kerr-scalar-self-force-and-wiggles}
observed and fitted $\ell\,{=}\,m\,{=}\,1$, $\ell\,{=}\,m\,{=}\,2$,
$\ell\,{=}\,m\,{=}\,3$, and $\ell\,{=}\,m\,{=}\,4$ wiggle (QNM) modes.
Their figures~8 and~9 show the $\ell\,{=}\,m\,{=}\,4$ wiggle as having
an amplitude approximately $10^9$~times smaller than the $\ell\,{=}\,m\,{=}\,1$
wiggle; this is only detectable by virtue of the high accuracy and
low numerical noise level of their frequency-domain code.
In contrast, for this configuration we observed wiggles for $m \,{=}\,1$
but not for $m \,{\ge}\, 2$; this is likely because even the $m \,{=}\,2$
wiggles are already too low in amplitude to be visually observable
in the original time series.

Existing astrophysical models of extreme mass ratio
binaries~\cite{Miller-etal:probing-stellar-dynamics-in-galactic-nuclei,
Hopman-Alexander-2005}
and observations of highly spinning black holes~\cite{Brenneman:2013oba}
suggest that it is quite reasonable to expect some fraction of EMRIs
to fall within the region of parameter space where wiggles are excited
with significant amplitude.  (Both the magnitude of this fraction
and the absolute numbers of such systems are still very uncertain.)

Given that the QNM excitations appear not just in the
local self-force, but also in the gravitational waveform, a natural
question is whether they could be experimentally observed by LISA
or other detectors.  While this is certainly possible in principle,
there are two considerations which make it less likely in practice.
Most importantly, the effect is quite weak in all but the most extreme
cases.  In most of the gravitational-field cases investigated here,
it was necessary to zoom~in
on plots in order to see the wiggles visually, reflecting the fact that
their magnitude represents at most a few percent of the total signal.
A second consideration in terms of detectability is that the dynamical
evolution of EMRIs may tend to avoid the wiggles region of parameter
space (e.g., if most EMRIs evolve to low orbital eccentricities
while still at relatively large periastron radii).  This would imply
that the event rate for \emph{detectable} EMRI wiggles would be
quite low.

Despite these concerns, it would be worthwhile to conduct
a more thorough study to quantitatively address the question of
detectability of QNM wiggles by LISA or future next-generation
gravitational-wave detectors.
It may even be the case that advanced data analysis
techniques could be used to boost the detectability. For example,
although an individual wiggle is weak, it will repeat for each orbit
throughout the entire inspiral. As noted
by~Ref.~\cite{Nasipak-Osburn-Evans-2019:Kerr-scalar-self-force-and-wiggles},
wiggles will appear with almost the \emph{same frequency} throughout
the inspiral (the QNM frequency only depends on the mass and spin
parameters of the larger black hole, and these change very little
during the inspiral).  Moreover, this frequency is much higher
than than the main orbital frequency, potentially making it easier
to separate these signal components in data analysis.

The analysis done here has been somewhat post-hoc, in that we first
identified a feature in the signal and then fit this feature to a
damped sinusoid representing a QNM ringdown. Our intuitive interpretation
of this QNM ringdown is that it is a result of strong QNM excitation near
periapsis, which is then encountered over an extended period later in the
orbit. The self-force in curved spacetimes arises from \emph{nonlocal}
interactions of the object with its self-field, which was generated in the
object's past and scattered off the spacetime curvature. The association of
wiggles with QNM excitations suggest that they represent a situation where
this nonlocal nature of the self-force is particularly apparent.
To more explicitly develop this interpretation, it may
be informative to attempt the analysis in the other direction, first
by starting with a model for a QNM excitation from a
burst of radiation generated near a periapsis passage, and then
comparing such a model to the observed signal. This approach would
allow one to pinpoint where in the orbit the QNM
excitation occurs, would give a deeper understanding of the effect,
and may even provide a link to geometric features such as caustics
and the propagation of waves on black hole spacetimes. A Green
function
approach~\cite{Casals:2013mpa,Wardell-etal-2014:self-force-via-Green-fn}
would be a natural choice for such a study, but is quite distinct
from the methodology used in this paper so we leave it for future
work.

In this work we have focused mostly on systems with somewhat realistic
Kerr spins $J/M^2 \lesssim 0.999$.  Initial investigations of the
near-extremal regime suggest a rich phenomenology, involving many
different QNMs at fixed $\ssl$ and $m$.  One puzzling result is the
$\ssl=m=2$ mode generated at $\Scri^{+}$ by a particle orbiting a black hole
with spin $J/M^2=0.999\,99$ and the same orbital frequencies as the
$(\tilde{a},p,e) = (0.99,3,0.8)$ orbit, shown in figure~\ref{fig:WTF}.
One of the
puzzling aspects of this waveform is that the decaying wiggles in
the highlighted area have a frequency between~$0.93 M^{-1}$ and~$0.98 M^{-1}$
(depending on where it is measured), while the nearest QNMs all have
frequency close to~$0.995 M^{-1}$.  Whether this is the result of some
complicated collective behaviour of the QNMs or some new physical effect
is currently unclear, and should be investigated in future works.
\footnote{
	 While we were making final revisions to this manuscript
	 Rifat, Khanna,
	 and Burko~\cite{Rifat-Khanna-Burko-2019:wiggles-in-near-extremal-Kerr}
	 reported a detailed study of wiggles in near-extremal
	 Kerr spacetimes, particularly the
	 $(\tilde{a},p,e) = (0.999\,99,2.918,0.807)$ system.
	 Their results are consistent with ours; they find
	 that the anomolous wiggle frequencies are an
	 intermediate-time effect caused by the superposition
	 of many simultaneously-excited Kerr QNMs.
	 }


\section{Acknowledgments}
\label{sect:ack}

We thank Leor Barack for invaluable discussions throughout the
course of this research.
We thank Richard~Brito for useful discussions about QNMs.
We also thank Dan Kennefick, Scott Hughes, Marc Casals, Peter Zimmerman,
Conor O'Toole and Adrian Ottewill for useful discussions on the results
of this paper.

JT thanks the Alexander von Humboldt Foundation for fellowship funding
for my stay at the Max-Planck-Institut f\"{u}r Gravitationsphysik
(Albert-Einstein-Institute),
and
the AEI (Division of Astrophysical and Cosmological Relativity)
and Indiana University (Office of the Vice Provost for Research,
Center for Spacetime Symmetries,
and Department of Astronomy)
for additional funding.

MvdM was supported by European Union's Horizon 2020 research and
innovation programme under grant agreement No.~705229.

Some of the numerical results in this paper were obtained using the
IRIDIS High Performance Computing Facility at the University of
Southampton and the Karst and Data Capacitor facilities at Indiana University
(supported by the U.S.~National Science Foundation under Grant No.~CNS-0521433,
by Lilly Endowment, Inc.{} through its support for the IU Pervasive
Technology Institute,
and by the Indiana Metabolomics and Cytomics (METACyt) Initiative).

\bibliography{journalshortnames,commongsf,jt-new,aei-references,meent,wardell}
\end{document}